\documentclass[]{tCPH2e}

\def\etal{\mbox{et al.}}
\def \slhclum{$10^{35}\ \mathrm{cm}^{-2}\mathrm{s}^{-1}$}
\def \lhclum{$10^{34}\ \mathrm{cm}^{-2}\mathrm{s}^{-1}$}
\def \lowlum{$10^{33}\ \mathrm{cm}^{-2}\mathrm{s}^{-1}$}
\def \timeslhclum#1{$#1\times 10^{34}\ \mathrm{cm}^{-2}\mathrm{s}^{-1}$}
\def \ifb {fb$^{-1}$}
\def \ipb {pb$^{-1}$}

\def \lambdahhh {\ifmmode \lambda_{\sss HHH}\else $\lambda_{\sss HHH}$\fi}
\def \lambdahhhsm {\ifmmode \lambda_{\sss HHH}^{\sss SM}\else 
$\lambda_{\sss HHH}^{\sss SM}$\fi}
\def \mh {\ifmmode m_H \else $m_H$\fi}
 
\def    \=              {\;=\;} 

\def    \lsim {\raisebox{-3pt}{$\>\stackrel{<}{\scriptstyle\sim}\>$}}  
\def    \gsim {\raisebox{-3pt}{$\>\stackrel{>}{\scriptstyle\sim}\>$}}

\newcommand     \be     {\begin{equation}} 
\newcommand     \ee     {\end{equation}} 
\newcommand     \ba     {\begin{eqnarray}} 
\newcommand     \ea     {\end{eqnarray}} 
 
\newcommand     \sss            {\scriptscriptstyle} 
 
\newcommand     \avg[1]         {\left\langle #1 \right\rangle} 
\newcommand     \ptmin     {\ifmmode p_{\scriptscriptstyle T}^{\sss min} \else 
                           $p_{\scriptscriptstyle T}^{\sss min}$ \fi}

\def \et   {\mbox{$E_{\scriptscriptstyle T}$}} 
\def \Emu  {\ifmmode{E_{\mu}
    }\else{$E_{\mu}$}\fi} 
\def \Enu  {\ifmmode{E_{\nu}
    }\else{$E_{\nu}$}\fi}

\def \nue  {\ifmmode{\nu_e}\else{$\nu_e$}\fi} 
\def \numu  {\ifmmode{\nu_{\mu}}\else{$\nu_{\mu}$}\fi}

\def \to   {\mbox{$\rightarrow$}}

\newcommand \jpsi{\ifmmode{J/\psi 
    }\else{$J/\psi$}\fi}

\begin{document}


\markboth{Michelangelo L. Mangano}{ }
\articletype{REVIEW}

\title{The super-LHC \\ {\small (To appear in Contemporary Physics)}}

\author{Michelangelo
  L. Mangano$^{a}$$^{\ast}$\thanks{
$^\ast$Email:
    michelangelo.mangano@cern.ch  \hfill {\large CERN-PH-TH/2009-164}}
\\\vspace{6pt}  $^{a}${\em{CERN, PH-TH, 1211 Geneva 23, Switzerland}};
\\\vspace{6pt}{} }

\maketitle

\begin{abstract}
We review here the prospects of a long-term upgrade programme for the
Large Hadron Collider (LHC), CERN laboratory's new proton-proton
collider. The super-LHC, which is currently under evaluation and
design, is expected to deliver of the order of ten times the
statistics of the LHC. In addition to a non-technical summary of the
principal physics arguments for the upgrade, I present a pedagogical
introduction to the technological challenges on the accelerator and
experimental fronts, and a review of the current status of the
planning.
\bigskip

\bigskip
\bigskip

\end{abstract}

\section{Introduction}
The Large Hadron Collider (LHC) is the new particle accelerator about
to start taking data at CERN's laboratory. It will collide protons
against each other, at a centre-of-mass energy of 14~TeV.  
Its primary
goal is to answer one of the deepest questions of physics today,
namely what is the origin of the elementary particles' masses. In
particular, it should be able
to verify whether the mechanism postulated by the current
theory of particle physics, the Standard
Model~\cite{Glashow:1961tr,Higgs:1964ia,'tHooft:1971fh,Kobayashi:1973fv},
is correct, or whether this requires additional ingredients. 

The
Standard Model, whose complete formulation dates back to the early
70's, has been shown over the past 30 years to accurately describe all
properties of the interactions among fundamental particles, namely
quarks, leptons and the gauge bosons transmitting the electroweak and
strong forces~\cite{Collaboration:2008ub,Erler:2008zz,Amsler:2008zzb}.  
Its internal
consistency, nevertheless, relies on a mechanism to break the symmetry
between electromagnetic and weak interactions, the so-called
electroweak symmetry breaking (EWSB). 

EWSB is a necessary condition
for elementary particles to acquire a mass. The reason is that weak
interactions have been shown experimentally, since the 50's, to be
{\it chiral}, namely to behave differently depending on whether the
projection of a particle spin along its momentum points towards the
direction of motion (positive chirality) or against (negative
chirality). Since the chirality of a massive particle can change sign
by changing Lorentz reference frame, the weak charge of a massive
particle cannot commute with the Hamiltonian, and the associated
symmetry must be broken. The simplest way to achieve this~\cite{Higgs:1964ia}
 is to assume the existence of a scalar field with a weak charge, 
the Higgs $H$, whose potential energy
is minimized with a non-vanishing value of its matrix element on the
vacuum state, $\avg{H}=v\ne 0$. This leads to spontaneous symmetry
breaking. The measured strength of the weak interactions and the mass
of their carriers, the $W^\pm$ gauge bosons, fix the value of $v\sim
247$~GeV, thus setting the natural mass scale for weak phenomena. 

The
fluctuations of the Higgs field around its vacuum state give rise to a
particle, the Higgs boson, whose mass $m_H$ is a free parameter of the
model. Direct searches at the LEP $e^+e^-$ collider have established a
lower limit $m_H \gsim 114$~GeV, important constraints in the mass
range around 170~GeV have been recently achieved by the Tevatron
experiments~\cite{cdfd0:2009pt}, and theoretical analyses of the
consistency of the model set an upper limit, around 800~GeV. The
design of the LHC collider and of its two largest experiments,
ATLAS~\cite{atlas} and CMS~\cite{cms}, has been tuned to enable the
full exploration of this mass range, searching for a broad
variety of the Higgs production and decay processes predicted by the
Standard Model.

The timeline for these searches is outlined in the left plot of
fig.~\ref{fig:H-susy-reach}, taken from~\cite{Blaising:2006qd}.  This
shows the amount of data, needed by each of the
two experiments to establish a $5~\sigma$ discovery, or a 95\%CL
exclusion, as a function of the Higgs mass. The present planning of
LHC operations foresees the delivery of a few~100~\ipb\ of data during
the first year, and of the order of 1--few~\ifb\ over the next couple
of years at a luminosity of about \lowlum. After reaching the {\it
  nominal} luminosity of \lhclum, the LHC should start delivering
about 60~\ifb per year, when accounting for down-time and running
efficiency.

A comparison with fig.~\ref{fig:H-susy-reach} therefore shows that,
within 2--3 years of data taking, the Standard Model Higgs boson will
be discovered, or entirely excluded, over the full mass range. In
either case, this will signal the beginning, rather than the
completion, of the LHC physics programme. Should the Higgs boson be
found, an extensive campaign of studies of its detailed properties
will be required, to confirm that they match the expectations of the
Standard Model, or to detect possibly minor deviations, unveiling a
framework for EWSB more elaborate than the minimal one postulated by
the Standard Model. If the Higgs boson is not found, a radical
departure from the Standard Model will be needed, and the searches to
understand what other mechanism is responsible for EWSB will begin.
\begin{figure}
\begin{center}
\begin{minipage}{150mm}
\hfil
\resizebox*{12cm}{!}{\includegraphics{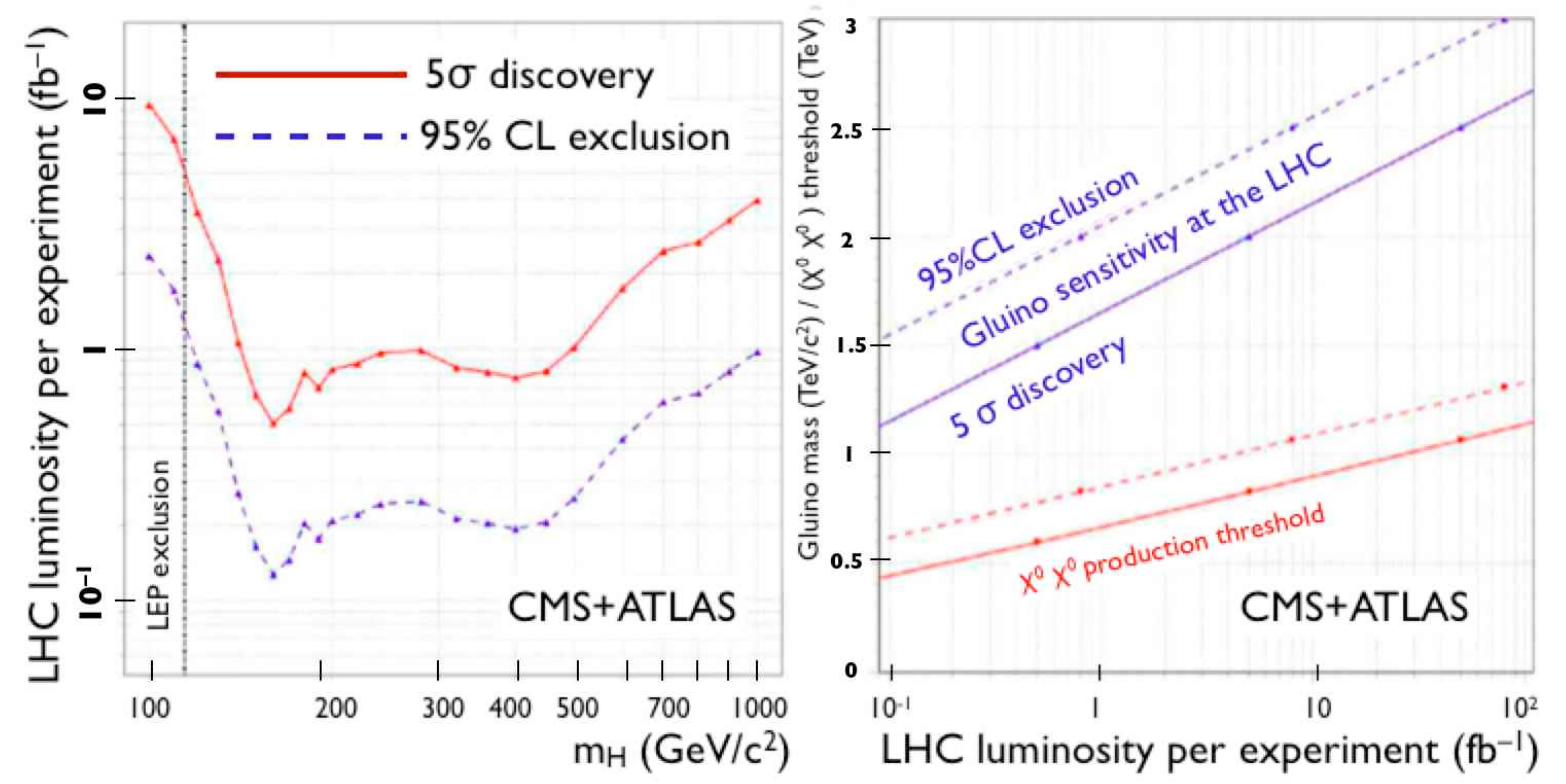} }
\hfil
\caption{Discovery reach and exclusion limits at the LHC for a
  Standard Model Higgs (left)
  and for gluinos in a supersymmetric theory (right), as a function of
  the integrated luminosity~\cite{Blaising:2006qd}.}
\label{fig:H-susy-reach}
\end{minipage}
\end{center}
\end{figure}

The LHC experiments are designed to be able to tackle these further
challenges, and the accelerator must be in the position to continue
delivering larger and larger amounts of data to allow them to pursue this
target. The goal of this review is to outline the potential of the LHC
to further push the study of EWSB and the search for other new
phenomena beyond the Standard Model (BSM), and to summarize the
technological challenges that this entails, both in the development of
a long-term higher-luminosity phase of the LHC accelerator, the
super-LHC (sLHC), and in the upgrade of the detectors, to allow them
to operate under the extreme experimental conditions that such a
higher luminosity will create. The first part of this review will
focus on the physics goals, discussing the possible measurements aimed
at more firmly establishing the nature of the EWSB and at probing
other new phenomena, and indicating the improvements that can be
obtained by extending the LHC operations to data samples a factor of
10 larger than what is currently foreseen by the base programme. A few
remarks will be included on the impact of an energy upgrade of the
LHC. The second part will discuss the possible evolution of the
accelerator complex to deliver 10 times the luminosity, including a
pedagogical overview of the main accelerator physics concepts required
to appreciate the challenge. A third part will address the
experimental constraints and the progress foreseen for the detectors.

A summary of the expectations for the first two years of measurements
at the LHC can be found in~\cite{Gianotti:2005fm}. The first complete
discussion of the physics potential of the sLHC has been presented
in~\cite{Gianotti:2002xx}.  Most of the results shown here are taken
from this document: while in the ensuing years new theoretical ideas
and new studies of the experimental prospects have appeared, the
examples discussed here well illustrate the potential of the LHC, the
main difficulties inherent in the analyses, and the added value
provided by the sLHC. The first technical assessment of the
feasibility of the accelerator upgrade is documented
in~\cite{Bruning:2002yh}. The discussion presented here is based on
the latest upgrade plans, and more details can be found
in~\cite{Scandale} and in~\cite{SLHC-AT}, an excellent series
of introductory lectures to be subject.

\section{The physics potential of the LHC programme}
As discussed in the introduction, the confirmation and exploration of
the mechanism driving the electroweak symmetry breaking is today the
main priority of particle physics. The Standard Model defines without
ambiguity the mechanism which brings the Higgs boson to
existence. There are nevertheless good reasons for theorists to
suspect that physics {\em beyond} the Standard Model should play a key
role in the dynamics of EWSB~\cite{Barbieri:2004ii}.  To start with,
one would like to have a framework within which the smallness of the
weak scale $v\sim247$~GeV, relative to the Planck scale $M_{Pl}\sim
10^{19}$~GeV, is the natural result of dynamics, as opposed to a
random accident. Furthermore, there are unequivocal experimental
indications that BSM phenomena are required to explain what is
observed in the universe: the Standard Model can explain neither
the existence of dark matter~\cite{Drees:2008zz}, 
nor the ratio of baryons and radiation
present in the universe. In addition, although neutrino masses could
be incorporated with a minimal and trivial adjustment of the Standard
Model spectrum, the most compelling explanations of how neutrinos
acquire such a small mass rely on the existence of new phenomena at
scales of the order of the Grand Unification, $M_{GUT}\sim
10^{15}$~GeV~\cite{Mohapatra:2005wg}. 
Furthermore, there are several questions that within the
Standard Model cannot be addressed, but that could acquire a
dynamical content in a broader framework. As an example, consider the
issue of what is the origin of the three quark and lepton generations
and of the diverse mass patterns between and within them. Since EWSB
is ultimately responsible for the generation of masses, with the
differentiation between flavours and the consequent appearance of
mixing angles and CP violation, speculating a relation between EWSB
and the flavour structure of the fundamental particles is unavoidable.
This relation is trivial in the context of the Standard Model, where
the flavour structure is determined by the couplings with the Higgs
field, which are free, arbitrary, parameters. By contrast, in most BSM models
the low-energy flavour structure emerges from a specific dynamics,
and relations between masses and mixings of different particles are in
principle calculable. 

EWSB therefore brings together the two main elements of the Standard
Model, the gauge and the flavour sectors. Their current theoretical
description has so far survived the most stringent experimental tests,
but both components are vulnerable, and liable to crack under the
weight of new data. The various BSM proposals anticipate new phenomena
to appear at the TeV mass scale: supersymmetry~\cite{Wess:1973kz,Kane:2000ns}, 
which implies the
existence of a new boson for each fermionic particle of the
Standard Model, and of a new fermion for each known boson; 
new forces, mediated by heavy gauge bosons, possibly
restoring at high energy the chiral left-right asymmetry of the
low-energy world; new strong interactions and new fermions,
responsible for the dynamical generation of EWSB; compactified space
dimensions, leading to the existence of an infinite Kaluza-Klein
  spectrum of particles corresponding to each known one, with ever increasing
and linearly-spaced masses; composite structures
within what are considered as fundamental, elementary particles; and
more, with a continuous emergence of new ideas and proposals to embed
the Standard Model into a more complete theory.  The LHC will be the
first accelerator operating at energies high enough to explore a large
fraction of these proposals.  This section will present a few examples
of the LHC discovery potential and of the benefits of a luminosity
upgrade.  They are just meant as illustrative, and a more detailed
account is documented in ref.~\cite{Gianotti:2002xx}.

In addition to ATLAS and CMS, four other approved experiments will contribute
to the completion of the LHC physics programme: ALICE~\cite{alice}, dedicated to
the study of relativistic heavy ion collisions; LHCb~\cite{lhcb},
dedicated to the study of of the properties of $b$-flavoured hadrons;
TOTEM~\cite{totem}, dedicated to the measurement of total and elastic cross
sections and LHCf~\cite{lhcf}, 
dedicated to the study of inclusive forward photon
and $\pi^0$ spectra. None of these experiments will be engaged in the
operations at the highest luminosities delivered by the sLHC, and this
review will primarily focus on the issues relevant to the physics
programme of ATLAS and CMS. 

\subsection{The Higgs sector and EWSB}
As shown in fig.~\ref{fig:H-susy-reach}, the discovery of the Standard
Model Higgs boson should be achievable by ATLAS and CMS independently
with luminosities in the range of 10~fb$^{-1}$ for the full mass range
$114<m_H\lsim 800$ GeV, and already with $\sim$1~fb$^{-1}$ in the
regions around 160 and 400~GeV, where the $H\to Z Z^{(*)} \to 4$
leptons decay mode provides a very clean signal. Following the
discovery, the main focus will become the quantitative study of the
Higgs properties. The goal will be to establish whether it behaves as
expected in the Standard Model, or whether BSM physics is
present. Contrary to the rather generic expectation that the LHC will
detect the Higgs, the issue of its precise nature is more open, as
different BSM theories make different predictions for the precise
nature of the EWSB mechanism. In some cases it will be straightforward
to estabish the incompatibility of the detected Higgs with the
Standard Model. For example, a value of $m_H \gg 200$~GeV would clash
with the result of electroweak fits.  Likewise, a production
cross-section significantly different from what is calculated would be
a sign that either the couplings are different than expected, or that
new states exist affecting the decay branching ratios, or both.  Even
a Higgs in the right mass range and with production rates roughly
compatible with the Standard Model may still hide some underlying BSM
dynamics~\cite{Giudice:2007fh}, which could become evident once more
accurate measurements become available. What is certain is that in
this phase there will be no limit to the need for accuracy and thus
for data statistics.

\begin{figure}
\begin{center}
\begin{minipage}{150mm}
\subfigure[]{
\resizebox*{7cm}{!}{\includegraphics{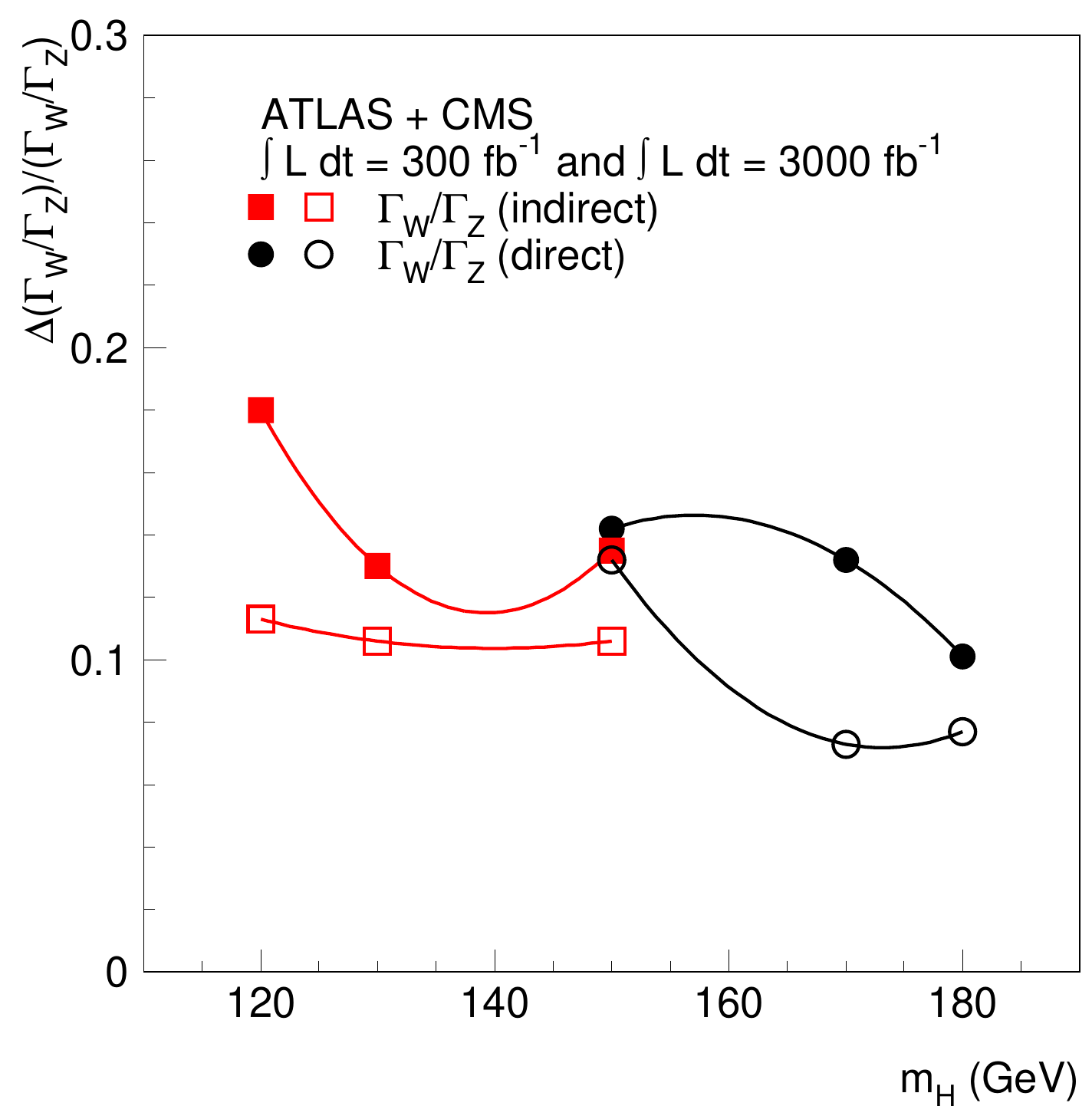}}}%
\hfil
\subfigure[]{
\resizebox*{7cm}{!}{\includegraphics{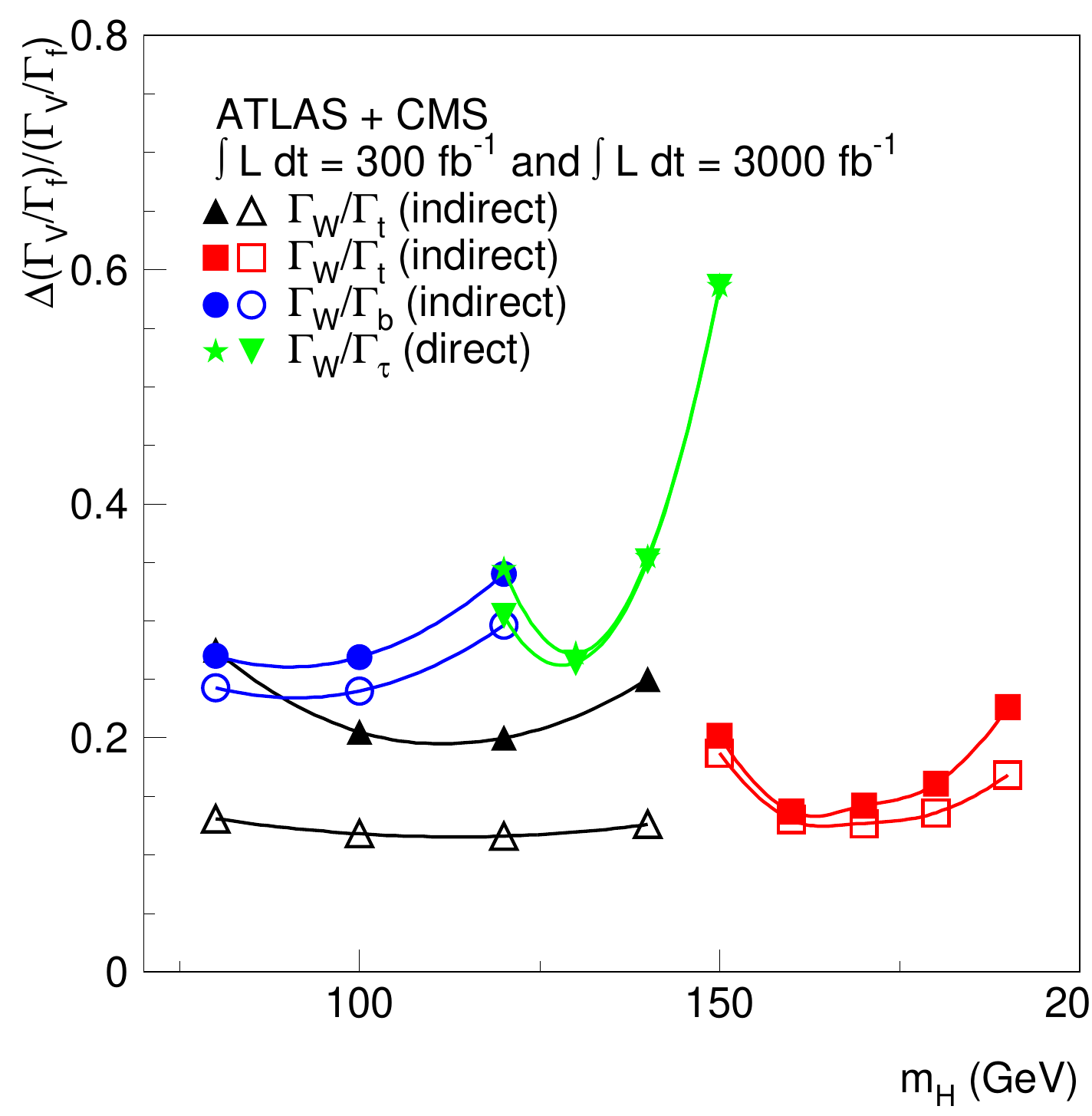}}}%
\caption{Expected uncertainties on the measured ratios of the 
 Higgs widths to final states involving  bosons only (a) and 
 bosons and fermions (b), 
 as a function of the Higgs mass. Closed symbols:  
 two experiments and 300~\ifb\  per experiment 
 (standard LHC); open symbols: two experiments and 
 3000~\ifb\  per experiment (sLHC). Direct and indirect 
 measurements have been included.}%
\label{fig:coupl}
\end{minipage}
\end{center}
\end{figure}
\subsubsection{Determination of Higgs couplings}
The expected precision in the determination of the Higgs couplings to
fermions and gauge bosons is shown in
fig.~\ref{fig:coupl}~\cite{Zeppenfeld:2000td,Gianotti:2002xx}. Model-indendent
results are given in terms of ratios of
couplings. These are less prone to theoretical or experimental
systematics, and are sufficient to exhibit possible deviations from the
Standard Model, which could be rather large in many BSM
scenarios, such as supersymmetry.
The limited improvement with the sLHC luminosity of some of these
measurements is due to the very conservative detector-performance
assumptions made in~\cite{Gianotti:2002xx}. This is to be reviewed in
the near future, with more realistic studies.

Higher luminosity will allow the measurements of decays that would
otherwise be too rare. An example is $H\to Z\gamma$,
which in the mass region 115-160~GeV is predicted to have a branching ratio of a
few~$\times 10^{-3}$. The expected significance for 600~fb$^{-1}$
(300~fb$^{-1}$ per experiment) is
$\sim 3.5~\sigma$~\cite{Gianotti:2002xx}, 
to become $\sim11~\sigma$ with a tenfold luminosity increase.
In the case of $H \rightarrow \mu^+\mu^-$, the Standard Model
branching ratio is of order $10^{-4}$ in the range 115-140~GeV, 
and the significance for
600~fb$^{-1}$ is below 3~$\sigma$, increasing to about 7~$\sigma$
at the sLHC~\cite{Han:2002gp}. 

\subsubsection{Observation of extended Higgs sectors}
Most BSM theories require the existence of more Higgs particles in
addition to the excitation of the field responsible for EWSB. 
For example, in supersymmetry there are two Higgs doublets, one
coupled to the up-type quarks, and one to the down-type quarks and to
charged leptons. After EWSB, 3 Higgs
fields emerge in addition to the Standard Model one, 
for a total of two CP-even ($h^0$ and $H^0$), one
CP-odd ($A^0$) and one charged ($H^\pm$). The spectrum and couplings
of these states are determined by the parameters $m_A$ (the mass of
the CP-odd Higgs) and $\tan\beta$ (the ratio of expectation values of
the neutral components of the two Higgs doublets).
The measurement of these masses and couplings provides
valuable information on the mechanism of supersymmetry breaking. The
LHC discovery potential for additional Higgs bosons in the minimal
supersymmetric standard Model (MSSM) is summarised in
  Fig.~\ref{fig:multiH}~\cite{Gianotti:2002xx}. Different
  shadings flag regions where different comibinations of Higgs bosons
  can be discovered. 
\begin{figure}
\begin{center}
\begin{minipage}{150mm}
\hfil
\resizebox*{9cm}{!}{\includegraphics{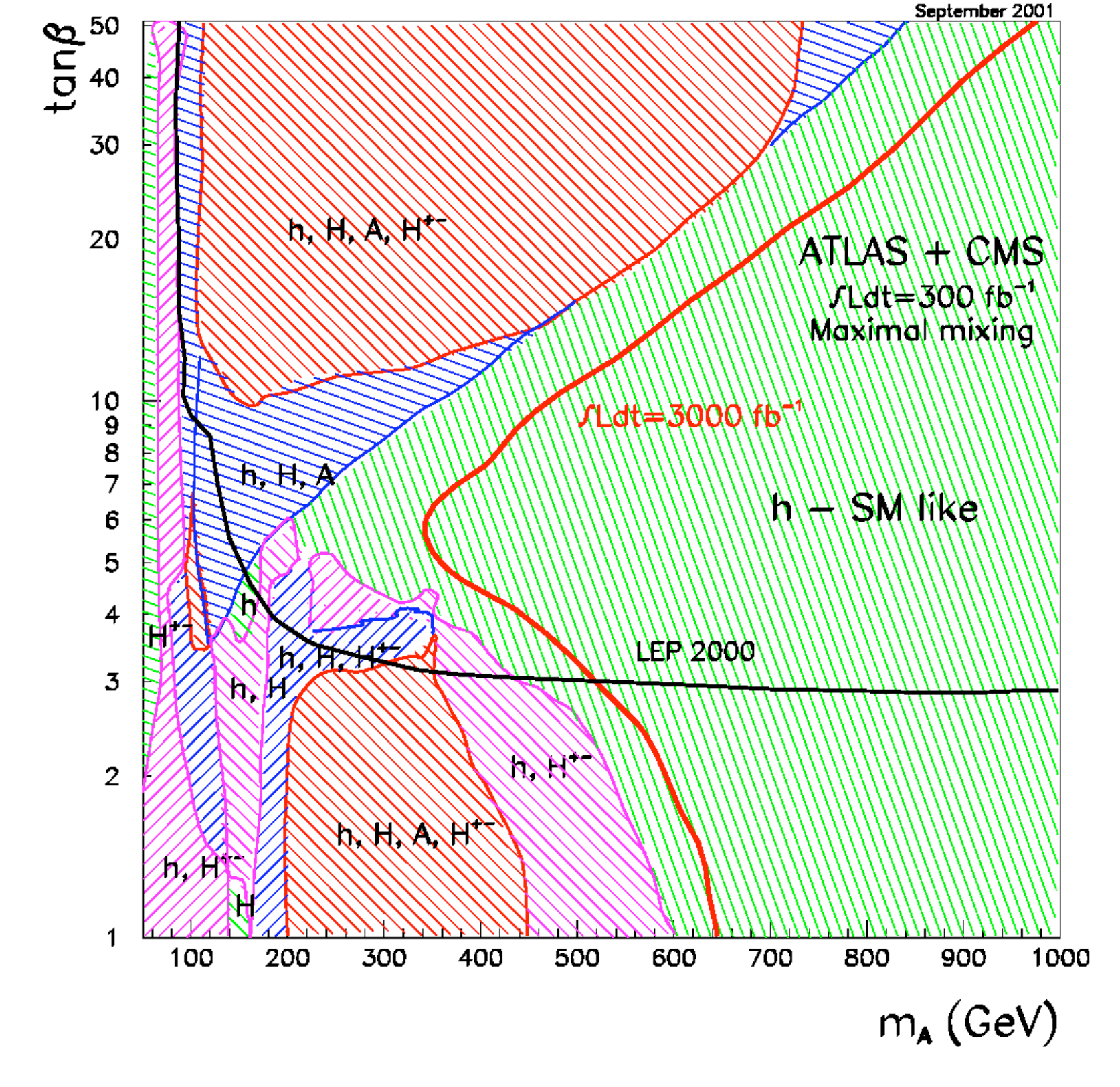}}%
\hfil
\caption{Regions of the MSSM parameter space where the various Higgs
  bosons can be discovered at $\geq5~\sigma$ at the LHC (for
  300~\ifb\ per experiment and both experiments combined) through
  their decays into Standard Model particles. 
In the region to the left of the rightmost contour at
  least two Higgs bosons can be discovered at the SLHC (for
  3000~\ifb\ per experiment and both experiments
  combined).\label{fig:multiH}}%
\end{minipage}
\end{center}
\end{figure}
This plot shows that over a good part of the parameter space the LHC
should be able to discover two or more Higgs bosons, except in the
region at large $m_A$ (the so-called ``decoupling limit"). In this
region, only the lightest Higgs boson $h$ can be observed, unless the
heavier Higgs bosons ($H,\ A,\ H^{\pm}$) have detectable decay modes
into supersymmetric particles. This means that the LHC cannot promise
a complete and model-independent observation of the heavy part of the
MSSM Higgs spectrum, although the observation of sparticles
(e.g. squarks and gluinos) will indicate that supersymmetry
exist, and tell implicitly that additional Higgs bosons should exist.
Figure~\ref{fig:multiH} also shows that the sLHC should be able to
extend significantly the region over which at least one heavy Higgs
boson can be discovered at $\geq~5~\sigma$ in addition to $h$
(rightmost contour in the plot).

\subsubsection{Strongly-coupled vector bosons}
\label{sec:strongew}
General
arguments imply that~\cite{Chanowitz:1985hj}, in absence of a Higgs
boson below $m_H\sim 1$~TeV, the scattering of
electroweak gauge bosons at high energy will show structure beyond
that expected in the Standard Model: resonances, or other phenomena,
must appear as a result of the strongly-interacting Higgs dynamics and
to correct the breakdown of unitarity of the
scattering amplitudes. Most recently, it has been pointed
out~\cite{Giudice:2007fh} that such phenomena could also occur if the
Higgs were light, should the Higgs be non-elementary, as in models
were EWSB is induced by a high-energy strongly-interacting sector.

In order to explore such signals it is necessary to measure final
states containing pairs of gauge bosons with invariant masses in the
TeV range and above.  The example given here refers to production of
$WZ$ pairs, in a chiral lagrangian model~\cite{Dobado:1999xb} in which
the unitarization of the scattering amplitude in the absence of a
Higgs boson is enforced by the presence of a massive vector
resonance. Figure~\ref{fig:WZ} shows the expected signal for a 1.5 TeV
resonance, at the LHC and at the sLHC, after applying very strict
analysis cuts to reduce the otherwise overwhelming QCD
backgrounds~\cite{Gianotti:2002xx}. At the LHC one requires the
presence of a forward and a backward jet (namely jets with
$\vert\eta\vert>2$\footnote{The pseudorapidity $\eta$ is related to
  the scattering angle $\theta$ by the relation
  $\eta=-\log\tan\theta/2$.}), with energies greater than 300 GeV, and
the absence of central ($\vert\eta\vert<2$) jets with transverse
energy $\et>50$~GeV.  At the sLHC, as discussed in
section~\ref{sec:physperf}, the high luminosity leads to a large
number of parasitic $pp$ collisions overlapping with the primary
interaction (the so-called pile-up events), and the associated
released energy would promote background events into the signal
region.  One is therefore forced to tighten these cuts to 400 and
70~GeV, respectively, thus reducing in parallel the signal efficiency.
The resonance is at the limit of the observability at the LHC, with
6.6~events of signal ($S$) expected over a background ($B$) of about
2.2~events around the region of the peak.  At the sLHC, on the other
hand, the signal has a significance of $S/\sqrt{B}\sim 10$. This
example underscores the importance of keeping the number of pile-up
events as small as possible, and of maintaining a good efficiency for
reconstruction of forward jets, two of the driving requirements for
the accelerator and the detector upgrades.
\begin{figure}
\begin{center}
\begin{minipage}{150mm}
\subfigure[]{
\resizebox*{7cm}{!}{\includegraphics{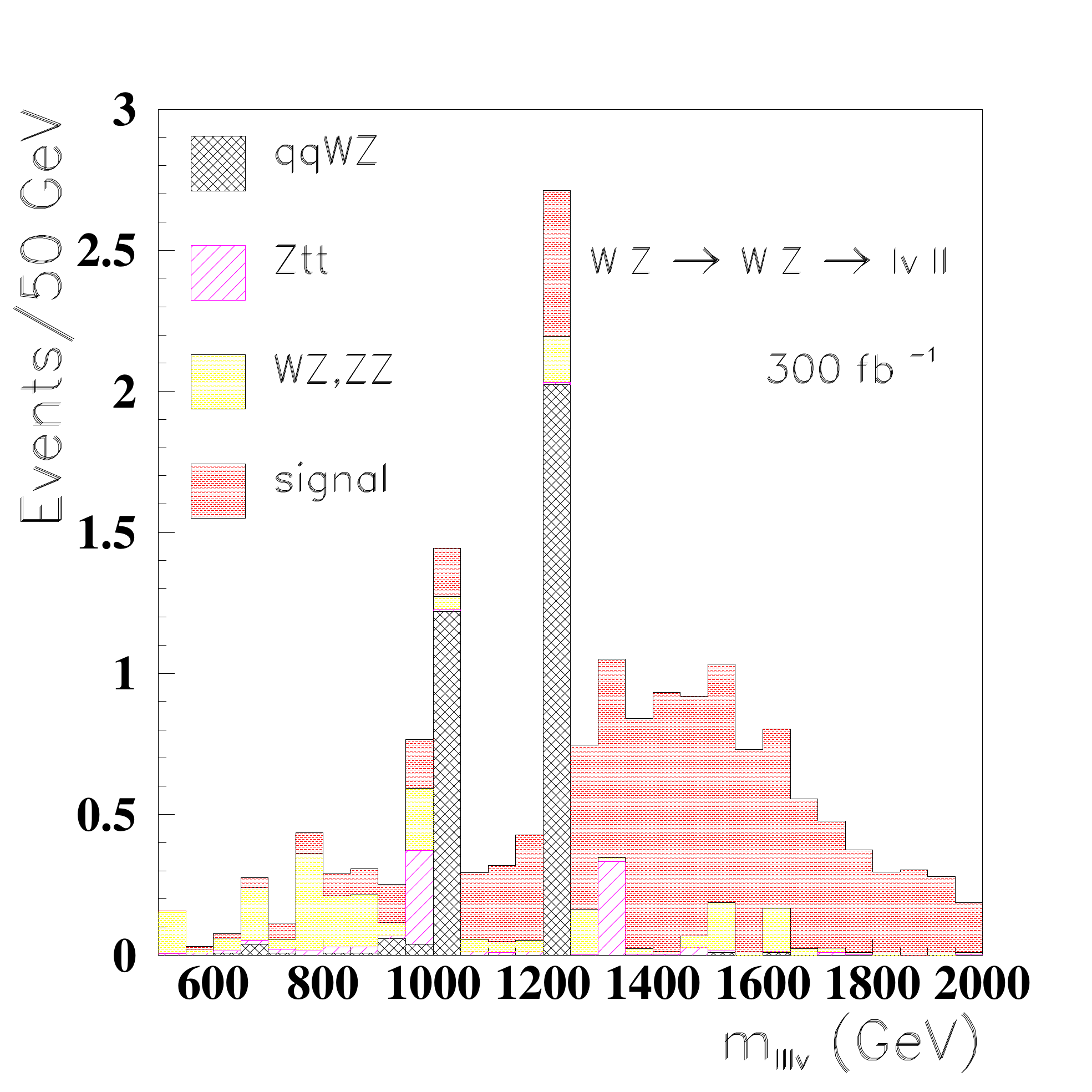}}}%
\hfil
\subfigure[]{
\resizebox*{7cm}{!}{\includegraphics{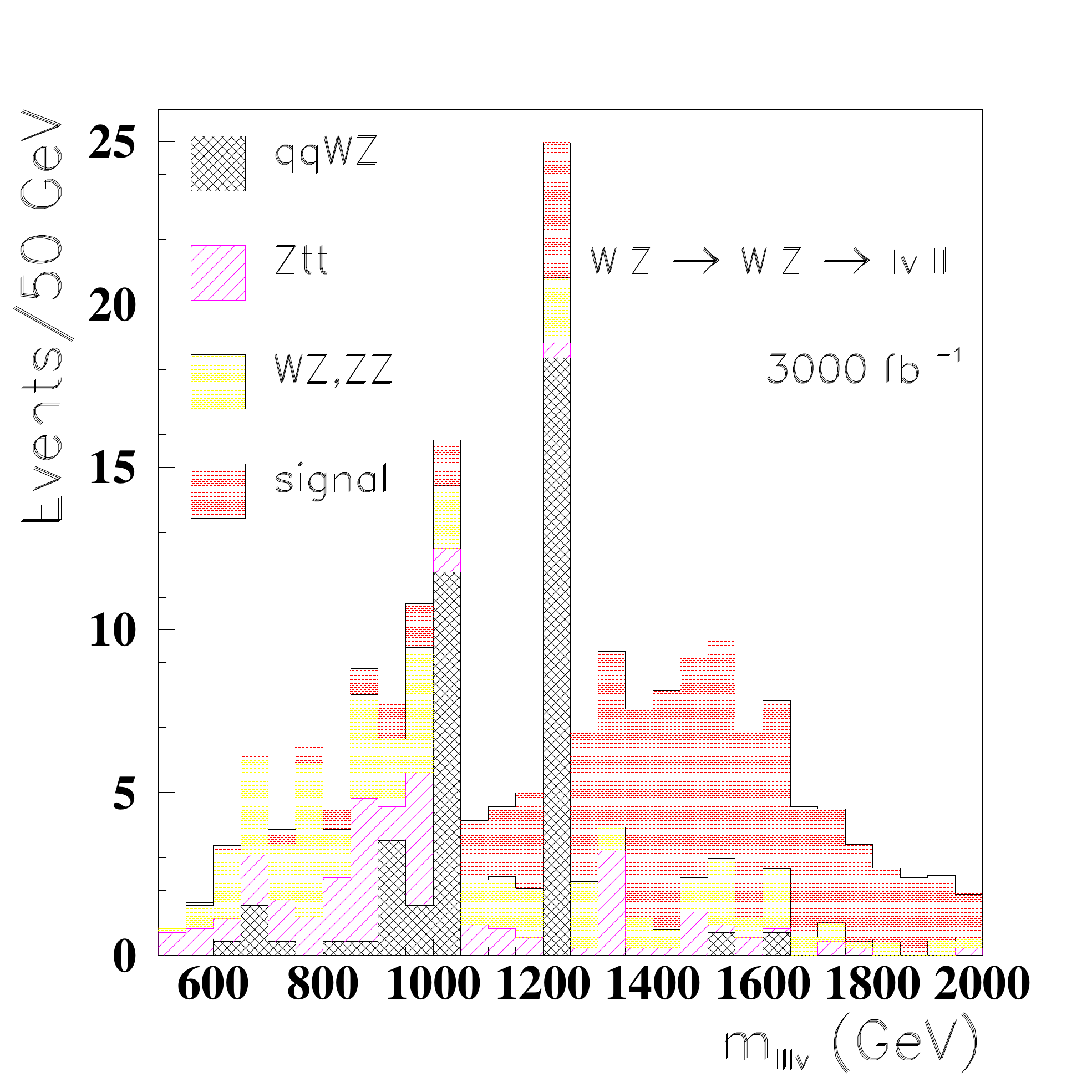}}}%
\caption{Expected signal and background for a 1.5 TeV WZ resonance 
in the leptonic decay channel for
300 fb$^{-1}$ (a) and 3000 fb$^{-1}$ (b).}%
\label{fig:WZ}
\end{minipage}
\end{center}
\end{figure}

\subsubsection{Gauge boson self-couplings}
A further crucial test of models for EWSB is the precise measurement of the
self-couplings of the electroweak gauge bosons. The Standard Model 
radiative corrections modify
the Born-level results at the level of per mille, setting
 the goal for precision measurements. 
\begin{table}
\tbl{Expected accuracies for the CP-conserving 
anomalous couplings of EW gauge bosons. The last column refers to an $e^+e^-$
linear collider (LC).
\label{tab:TGC}}
{\begin{tabular}{@{}lccccc}\toprule
Coupling  &    14 TeV & 14~TeV & 28~TeV & 28~TeV & LC (500~GeV) \\
          & 100~\ifb & 1000~\ifb & 100~\ifb & 1000~\ifb & 500~\ifb
\\
\colrule
$\lambda_\gamma$&0.0014&0.0006&0.0008&0.0002& 0.0014 \\
\colrule
$\lambda_Z$&0.0028&0.0018&0.0023&0.009 & 0.0013 \\
\colrule
$\Delta \kappa_{\gamma}$&0.034&0.020&0.027&0.013 & 0.0010 \\
\colrule
$\Delta \kappa_Z$&0.040&0.034&0.036&0.013 & 0.0016 \\
\colrule
$g_1^Z$&0.0038&0.0024&0.0023&0.0007 & 0.0050 \\
\botrule
\end{tabular}}
\end{table}
Table~\ref{tab:TGC}~\cite{Gianotti:2002xx} shows, for the
CP-conserving anomalous couplings of the electroweak (EW) bosons, the
accuracy expected with different options of luminosity and energy for
the LHC, as well as with a low-energy linear
collider~\cite{Weiglein:2004hn}. Notice that here
 the tenfold increase in luminosity is more powerful than a
twofold increase in energy at constant luminosity. The reason is that
the growth in the production rates for pairs of gauge bosons is only
logarithmic with beam energy, due to their rather low mass in relation
to the available phase-space, and higher luminosity is therefore
statistically more effective than higher energy.

\subsection{Supersymmetry}
\begin{figure}
\begin{center}
\begin{minipage}{15cm}
\hfil
\resizebox*{10cm}{!}{\includegraphics{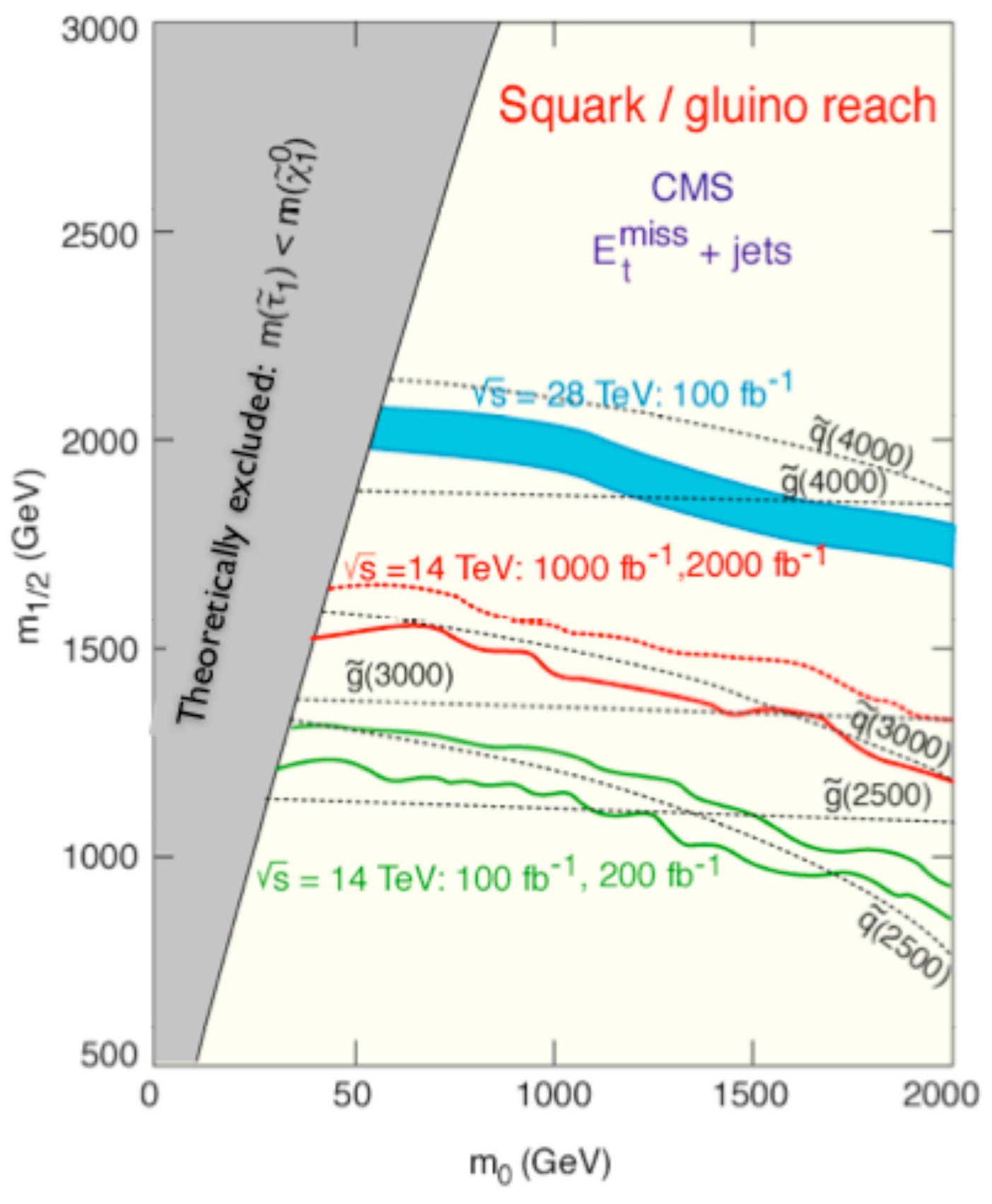}} 
\hfil
\caption{Expected $5~\sigma$ discovery contours for gluinos and
  squarks, as a function of the
  mass parameters $m_0$ and
 $m_{1/2}$. The various curves show
 the potential of the CMS experiment at the standard LHC (for luminosities of
 100~\ifb and 200~\ifb), at the sLHC 
 (for 1000~\ifb and 2000~\ifb), and at the DLHC
 ($pp$ collisions at double-LHC energy, $\sqrt{S}=28$~TeV).
\label{fig:susy}} 
\end{minipage}
\end{center}
\end{figure}
The current mass limits on supersymmetric particles reach 300--400
GeV, from the searches at the Tevatron
collider~\cite{Amsler:2008zzb}. As shown in
fig.~\ref{fig:H-susy-reach}, few~100~\ipb\ will be sufficient at the
LHC to extend the 5~$\sigma$ sensitivity up to well above the TeV
scale! The figure also shows that the mass sensitivity grows
logarithmically with the statistics, with each new decade in
integrated luminosity increasing the discovery reach by about
500~GeV. This trend will hold up to the sLHC luminosity, corresponding
to the mass reach of about 3~TeV shown in
fig.~\ref{fig:susy}\footnote{Above this mass threshold, the production
  cross section starts falling much more rapidly, due to the vanishing
  probability of finding quarks and antiquarks inside the proton
  carrying enough energy for the creation of such massive final
  states.}.  The discovery reach is shown here, for a specific
supersymmetric model, by the wiggly lines labeling the energy and
luminosity of the accelerator configuration. The model parameters on
the plot's axis, $m_0$ and $m_{1/2}$, determine the mass of squarks
and gluinos across the plane, as shown by the various equal-mass
contours.

At high mass, the search for gluinos and squarks and the study of
their final states will not be limited
by detector systematics, because of the large amounts of energy
released, and statistics will be the dominant limitation in these
studies.
But even with an early discovery of supersymmetry, with masses in the
1--2~TeV range, the extensive programme
of measurements that this will trigger (sparticle masses and
couplings) will benefit from the higher  sLHC statistics, as discussed
in detail in~\cite{Gianotti:2002xx}. In this case, however, the lower
energies of the jets and of the various final-state objects (leptons,
missing transverse energy, $b$-jets, etc) will be much more sensitive
to the high-luminosity environment of the sLHC, and maintaining
excellent detector performance will be the primary concern, as
discussed in section~\ref{sec:physperf}.

\subsection{New forces}
\begin{figure}
\begin{center}
\begin{minipage}{15cm}
\hfil
\resizebox*{8cm}{!}{\includegraphics{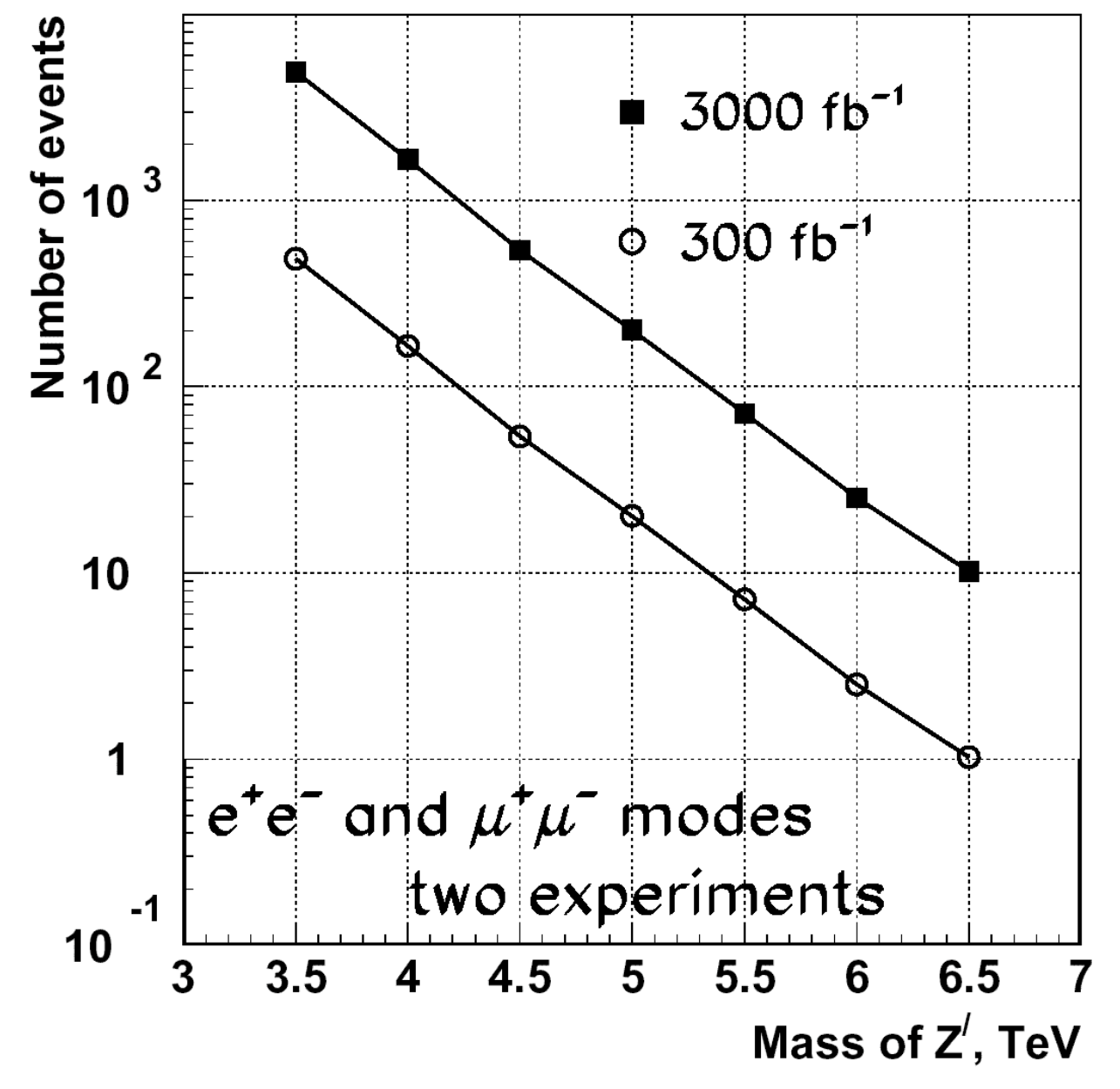} }
\hfil
\caption{Event rates for a $Z'$ with Standard Model couplings, at
  the LHC and sLHC.}
\label{fig:Zpreach}
\end{minipage}
\end{center}
\end{figure}
Figure~\ref{fig:Zpreach} shows the discovery reach for a new gauge boson
$Z'$, with Standard Model-like couplings, 
as a function of its mass. 
With 10 events to claim discovery, the
reach improves from ~5.3 TeV (LHC, 600~\ifb)
to ~6.5 TeV (sLHC, 6000~\ifb)       
\begin{figure}[tb]
\begin{center}
\begin{minipage}{15cm}
\hfil
\resizebox*{14cm}{!}{\includegraphics{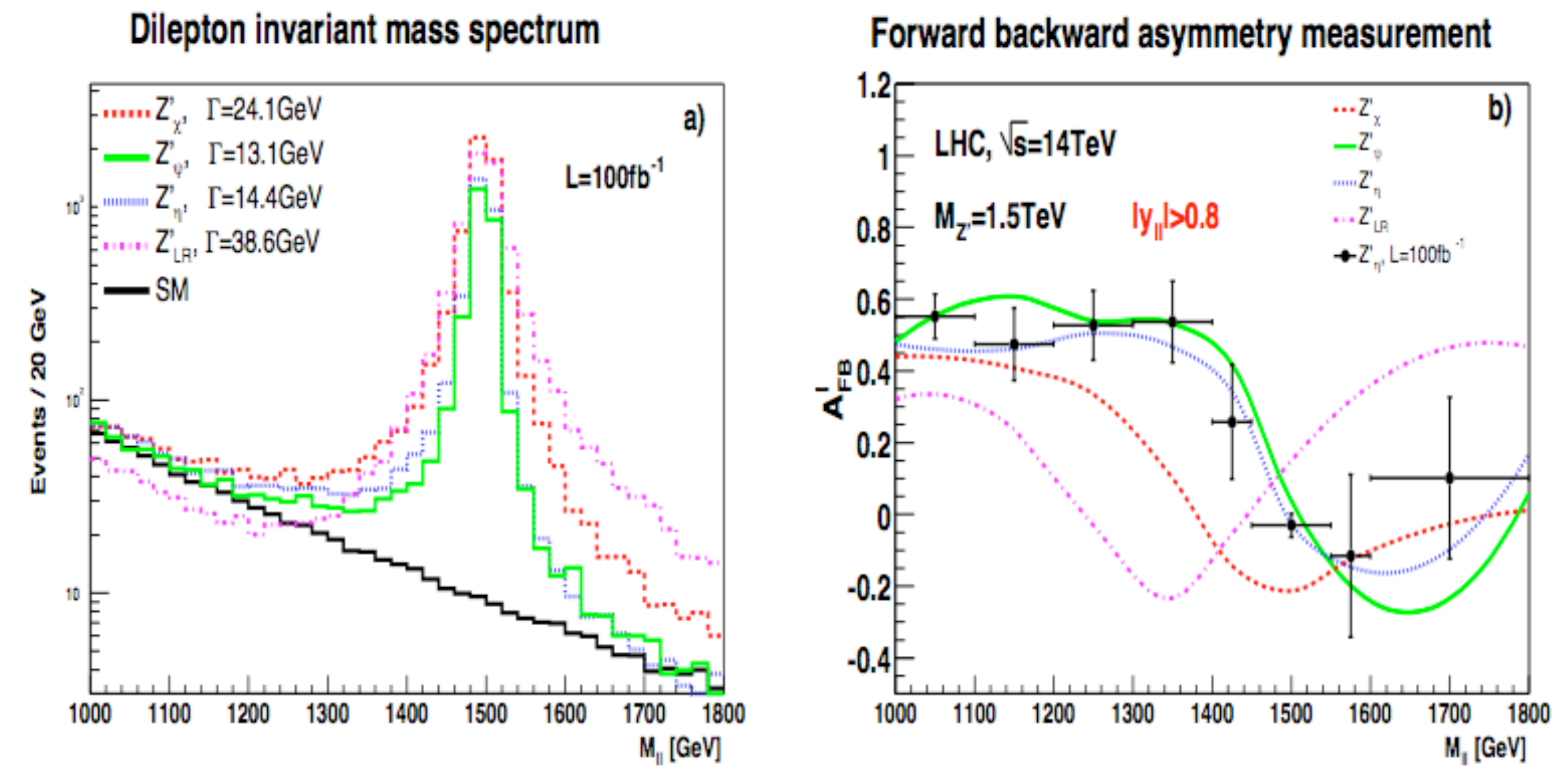} }
\hfil
\caption{For 
four models of $Z'$ bosons~\cite{Dittmar:2003ir}, we show:
(a) the spectrum of the dilepton invariant mass
  ($M(\ell\ell)$) and (b) the
  forward-backward asymmetry AFB as a
  function of $M(\ell\ell)$.}
\label{fig:Zpmodels}
\end{minipage}
\end{center}
\end{figure}
Notice that, in spite of the great discovery reach, the ability to
extract the values of couplings and to identify the specific nature of
the new force is confined to much lower masses, due to the limited
statistics.  This is shown in fig.~\ref{fig:Zpmodels}, in the case of
different models for a $Z'$ of 1.5 TeV~\cite{Dittmar:2003ir}, for $Z'$
decays to $\mu^+\mu^-$ pairs.  We
show the line shape of the dilepton mass distribution $M(\ell\ell)$, 
and the leptonic forward-backward asymmetry $A_{FB}$, as a
function of the dimuon invariant mass. Different curves correspond to
different $Z'$ couplings as obtained in various potentially
interesting models of new gauge symmetries. The points and relative
error bars correspond to a specific model, and to the expected
statistical uncertainty with an integrated luminosity of 100~\ifb. It
was shown in~\cite{Dittmar:2003ir} that a clear separation among
models can only be achieved for masses up to about 2.5~TeV.  A factor
of 10 increase in luminosity would extend this reach up to about
3.5~TeV, a mass beyond the reach on any future collider under
consideration today.

It is worth adding one more remark. It is unlikely that a $Z'$ will be
the only manifestation of new physics. Additional $Z'$s occur in most
GUT theories, providing, among other things, non-trivial interactions
to the otherwise sterile right-handed components of neutrinos. It is
well known that the gauge coupling unification predicted by GUT
theories is best verified in presence of supersymmetry. The
co-existence of supersymmetric particles and of a $Z'$ is therefore a
natural conjecture. While the LHC is a powerful machine to produce and
detect strongly-coupled supersymmetric particles, the production rates
of weakly-couple states (such as the sleptons, namely the scalar
partners of leptons) are rather small, and the backgrounds to their
detection are very large. Should these states lie below the threshold
for the direct decay of a $Z'$, their production through the $Z'$
resonance would greatly increase their observability. The study
in~\cite{Baumgart:2006pa}, for example, shows that the discovery reach
for sleptons would increase from 170--300~GeV, without a $Z'$, to over
1~TeV. The $Z'$ mass peak would provide a reference candle for the energy
of its decay products, making it possible to accurately determine
their mass, even in presence in their decay chain of undetected
particles, such as a neutralino, 
the lightest supersymmetric particle, and a dark matter candidate. 
In this case, one could directly
measure  the neutralino mass,
and see whether it is compatible with the properties of
dark matter. Statistics would be the main constrain to pursue a
complete study of the spectrum of new particles lying below the $Z'$.
The sLHC could thus become a $Z'$ factory, and acquire
many of the advantages so far attributed only to lepton colliders. 

\subsection{New structure}
A tenfold increase in the LHC luminosity should give access to jets
of up to $E_T\sim 4.5$~TeV (see Fig.~\ref{fig:jets}), thereby
extending the machine kinematic reach for QCD studies by up to 1~TeV.
\begin{figure}
\begin{center}
\begin{minipage}{150mm}
\hfil
\resizebox*{10cm}{!}{\includegraphics{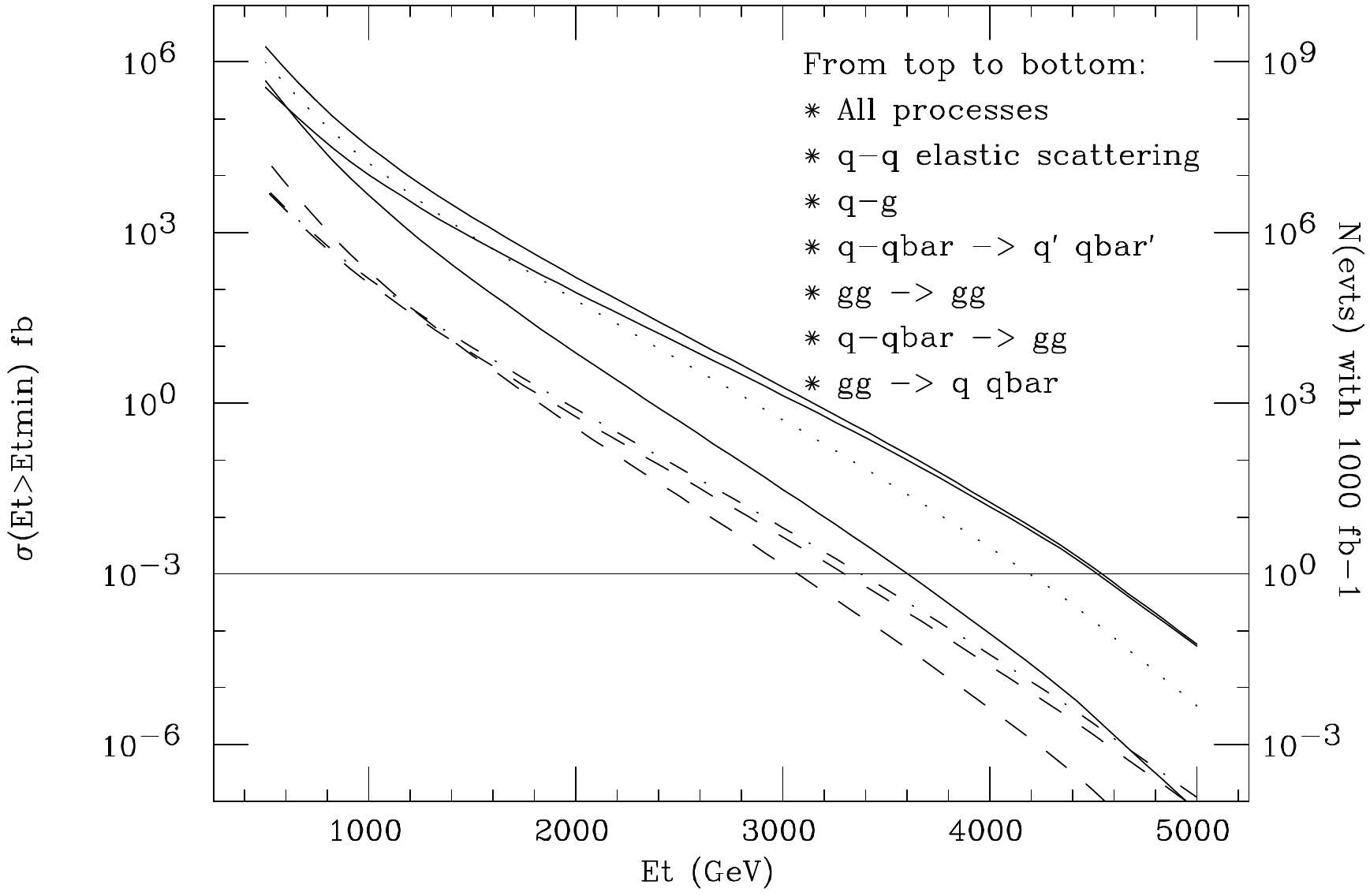} }
\hfil
\caption{Integrated production cross-section and 
 rates for inclusive central ($\vert \eta
  \vert < 2.5$) jets. The different curves label the 
  contributions of the various initial-state combinations 
  to the total cross-section.}
 \label{fig:jets}
\end{minipage}
\end{center}
\end{figure}
 This improved sensitivity should have an impact also on the search for
quark sub-structures.  Indeed, signals for quark compositeness should
reveal themselves in deviations of the high energy part of the jet
cross-section from the QCD expectation. The angular distribution of
di-jet pairs of large invariant mass provides an independent signature
and is less sensitive to systematic effects like possible
non-linearities in the calorimeter response. 
The compositeness scales that can be probed in this way at the LHC and
sLHC are summarised in Table~\ref{tab:compo}. For comparison, the
potential of a 28~TeV machine is also shown. It can be seen that a tenfold
luminosity increase would have an important impact for this physics,
comparable to the energy doubling.
\begin{table}
\begin{center}
\caption{The 95\% C.L. lower limits
  that can be obtained on the compositeness scale $\Lambda$
  by using di-jet angular distributions and for various
  energy/luminosity scenarios~~\cite{Gianotti:2005fm}. }
\label{tab:compo}
\vspace*{0.1cm}  
{\begin{tabular}{@{}lccccc}\toprule
Scenario         &    14 TeV 300 fb$^{-1}$&  14 TeV 3000 fb$^{-1}$
&28 TeV 300 fb$^{-1}$ &28 TeV 3000 fb$^{-1}$ \\
\colrule
$\Lambda$(TeV)  &     40  &     60 &     60   &    85 \\
\botrule
\end{tabular}}
\end{center}
\end{table}
  As these measurements involve only the calorimeters and jets in the
TeV range, they can be performed at the sLHC without major detector
upgrades.  Ability to extend the heavy-flavour tagging to the very
high \et\ region could however help disentangling the flavour
composition of a possible cross-section excess. Only a
fraction smaller than few~\% of the jets with $\et>2$~TeV should contain
bottom or charm quarks, therefore any indication of a long lifetime
component in these jets beyond this level would signal the presence of
new physics.

\section{The evolution of the accelerator complex}
The LHC injector chain is shown in fig.~\ref{fig:cerncomplex}. The
first stage of the acceleration takes place in the Linac2, a linear
accelerator with an output proton energy of 50~MeV. The proton-booster
synchrotron (PSB) increases the energy to 1.4~GeV, injecting into the
50-years old proton synchrotron (PS). This accelerates the beam to
26~GeV, and injects into the super proton synchrotron (SPS), out of
which 450~GeV protons are eventually injected into the LHC for the
start of the ramp up to the nominal energy of 7~TeV.
\begin{figure}
\begin{center}
\resizebox*{14cm}{!}{\includegraphics{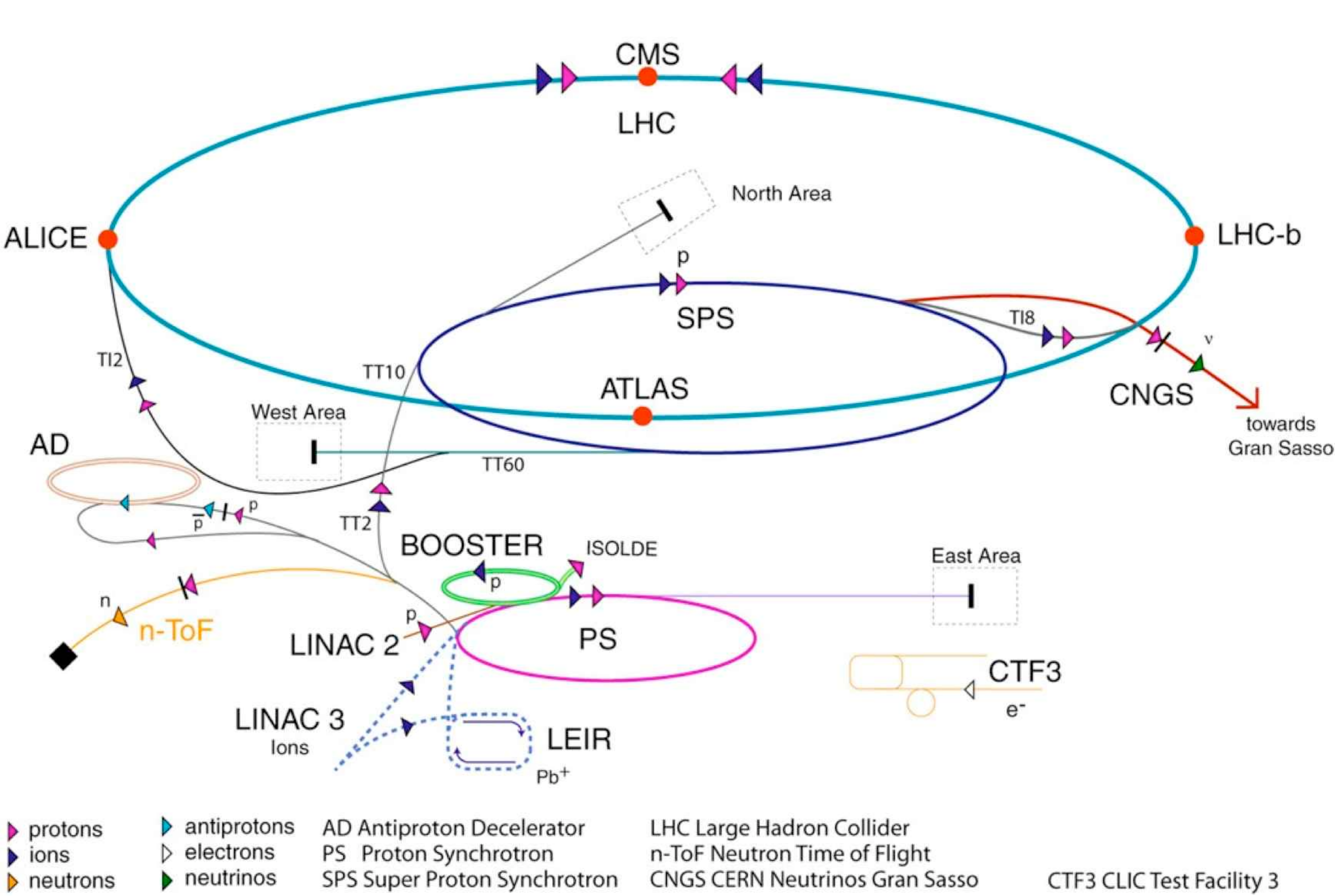}}%
\caption{Layout of the full CERN accelerator complex, including all
  elements of the LHC injector chain. The four interaction regions
  hosting the main LHC experiments, ALICE, ATLAS, CMS and LHCb, are
  also shown. }%
\label{fig:cerncomplex}
\end{center}
\end{figure}
To describe the fundamental constraints and evaluate the technological
options available for the luminosity upgrade, it is useful to briefly
summarize first some basic notations and notions of accelerator physics.

\subsection{The parameters controlling the collider luminosity}
As the protons travel around the ring, a sequence of pairs of
opposite-polarity (focussing and defocussing) quadrupole magnets
prevents the beam from blowing up and confines it within the
beampipe. During one full turn, an individual proton oscillates around
the ideal circular trajectory a number of times ({\it betatron
  oscillations}). To avoid the build up of resonance, this number, called
the {\it tune} ($Q$), should not be an integer. 
Any spread or shift of the tune ($\Delta Q$), due for example to
the beam-beam interactions as beams cross each other during the collision, 
should be kept small to avoid
hitting these resonances. The transverse size of the envelope of the
various trajectories, at a given point $s$ along the ring, is measured
by the betatron function $\beta(s)$, a quantity that, point by point,
depends on the local optics. 
Liouville theorem, on the other hand, constrains the possible
phase-space evolution of the beam. The associated Liouville invariant
is called the beam {\it emittance}, $\epsilon$. At a fixed momentum,
and in an ideal loss-less beam-transport scenario, this is a constant 
along the ring, while the so-called {\it normalized} emittance,
$\epsilon_n=\beta \gamma \epsilon$ (where $\beta=v/c$ and
$\gamma=1/\sqrt{1-\beta^2}$), is independent of momentum and is a
constant across the full beam acceleration and
storage path. Its value is defined at the earliest stage of the
acceleration process, and will be inherited, with some unavoidable
degradation, by the high-energy
components of the accelerator chain. 
The betatron function and the emittance combine to give the 
physical transverse size $\sigma$ of the beam at a point $s$:
$\sigma^2 \sim \epsilon \beta(s)$. This naturally leads to the following
relation for the peak collider luminosity:
\be \label{eq:lum}
L=\frac{f_r\, \gamma}{4 \pi} \; \frac{N_b^2 \, n_b}{\epsilon_n \,
  \beta^*} \; F \; .
\ee
Here $f_r$ is the revolution frequency, $N_b$ is the number of protons
per bunch, $n_b$ is the number of bunches, $\beta^*$ is the value of
the betatron function at the interaction point (IP), and $F<1$ is a factor
measuring the geometric loss of overlap between two bunches as they
cross at a given crossing angle. The luminosity can therefore be
increased by increasing the bunch current ($N_b$), the number of
bunches ($n_b$) and the geometric overlap ($F$), or by reducing
emittance or $\beta^*$. 

\subsubsection{Beam brightness}
As mentioned earlier, the normalized emittance is an invariant through
the full injector chain. Should the reduced phase-space acceptance of some
accelerator element create a
bottleneck, the emittance would increase, leading to a
degradation of the beam {\it brightness}~$\propto N_b/\epsilon_n$ and
of the peak luminosity. At this time, the LHC brightness is limited by
the characteristics of the Linac2, of the PSB and of the
PS. While these have been been shown to provide
the brightness required to reach the nominal LHC luminosity of 
\lhclum, any further luminosity increase based on a boost of the
brightness will require an upgrade of these low-energy elements of the
injector complex, as will be discussed later. 
\begin{figure}
\begin{center}
\resizebox*{14cm}{!}{\includegraphics{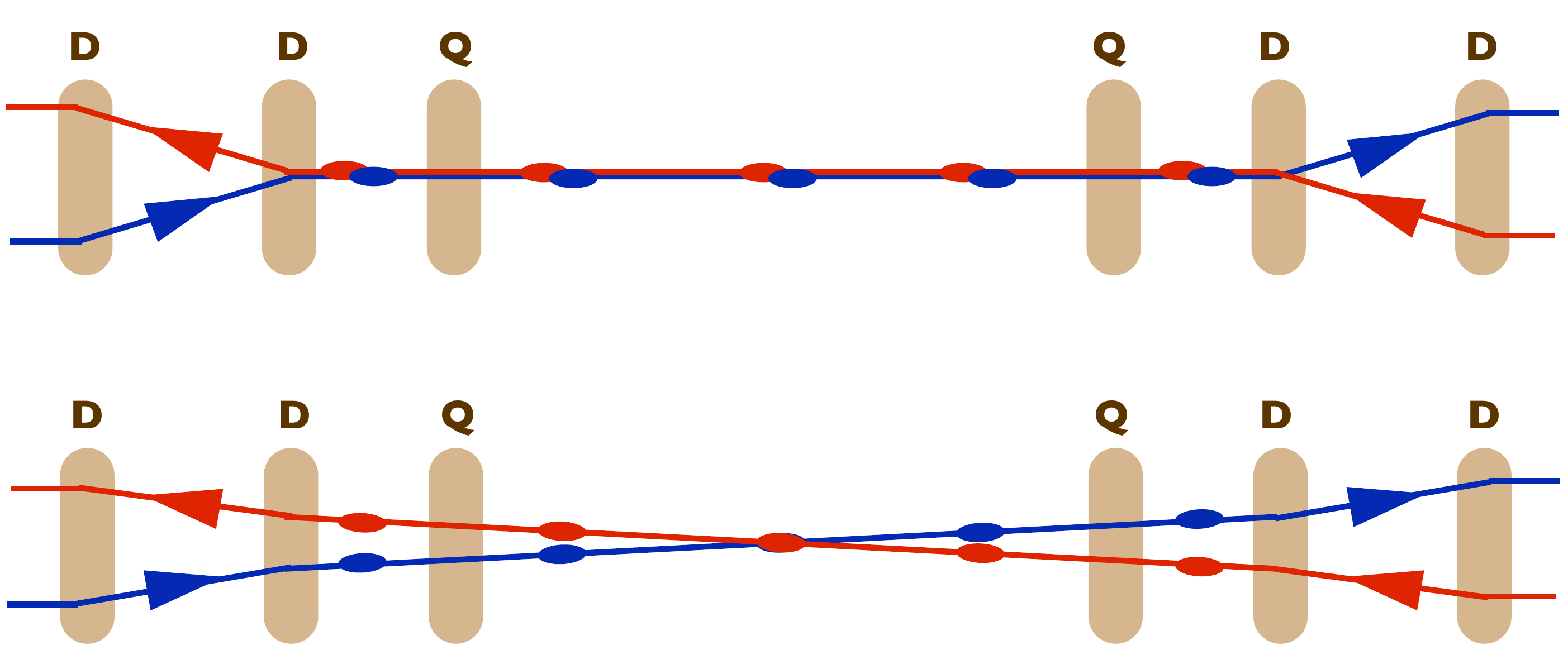}}%
\caption{The interaction region at the LHC, with (lower) and without
  (upper) a crossing angle.}%
\label{fig:crossingangle}
\end{center}
\end{figure}
An increase in brightness, on the other hand, leads to an increase in
the tune spread due to beam-beam interactions, since $\Delta Q\propto
N_b/\epsilon$ for a single head-on collision. 
To contain the tune shift from multiple collisions due to the short
bunch spacing, the present design of the
LHC requires collisions to take place at a non-zero crossing
angle, as shown in fig.~\ref{fig:crossingangle}. The upper
configuration in the figure shows a collision at zero crossing angle. Two dipole
magnets on each side of the interaction region (IR) bring the beams
collinear from their parallel but separated paths in the respective
beampipes; the bunch separation is 7.5~m, and thus each bunch will
cross about 30 opposite bunches during the 60~m trajectory
across the IR, before being redeflected back into its normal orbit,
with a 30-fold increase of the tune shift. To reduce this effect, weak
orbit correctors 
can help bring beams into collision with a small relative angle, as shown in the
lower configuration; the beam-beam interactions are suppressed,
due to their separation while away from the IP. 
In this configuration, however, the geometric overlap of the two
bunches at the IP is reduced, as shown in
fig.~\ref{fig:crabcrossing}(a),
 by a factor $F(\phi)=1/\sqrt(1+\phi^2)$,
where $\phi=\theta_c\sigma_z/(2\sigma_x)$ is the so-called Piwinski
angle. Here $\theta_c/2$ is the crossing angle w.r.t. to the horizontal, and
$\sigma_{z,x}$ are the longitudinal and transverse bunch
dimensions. There is therefore a competition between the need to
minimize the tune shift, which approximately decreases with crossing
angle by the factor $F(\phi)$, and the desire to increase the bunch
intensity, or reduce its transverse size $\sigma_x$. This highlights one
of the many constraints on the luminosity increase.
For nominal LHC operations, the crossing angle is 
$\theta_c=285~\mu$m and $F\sim 0.8$. 
A brilliant solution to this problem could come
from the development of {\it crab} RF
cavities~\cite{Palmer:1988yu,Oide:1989qz}, whose role is to tilt the
bunches before they enter the IR, ensuring their total overlap when
they cross, as shown in fig.~\ref{fig:crabcrossing}(b). So far, crab
cavities have only been tested in the KEK $e^+e^-$ collider, and a
vigorous R\&D is required to develop the technology for the
LHC~\cite{tuckmantelAT}. 
\begin{figure}
\begin{center}
\resizebox*{14cm}{!}{\includegraphics{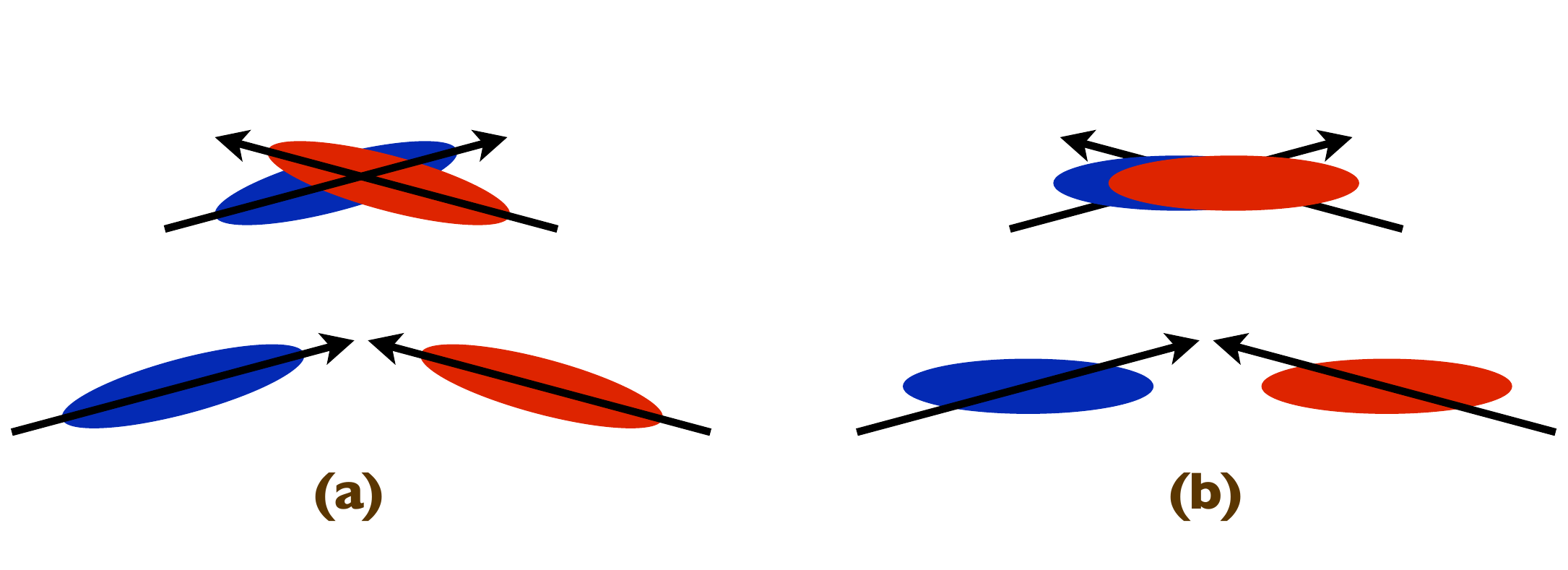}}%
\caption{Beam-beam overlap at the IP, in a normal configuration (a)
  and with a crab crossing (b).}%
\label{fig:crabcrossing}
\end{center}
\end{figure}

\subsubsection{The number of bunches}
The luminosity grows linearly with the number of bunches. The nominal
LHC will operate with 2808 bunches per beam, divided into bunch-trains
with a bunch-bunch spacing of
25~ns. Increasing the number of bunches while keeping $N_b$ fixed has
the great advantage for the experiments of maintaining the same number
of interactions within each bunch crossing, thus not increasing the
complexity (number of tracks, occupancy in the tracking chambers,
energy deposits in the calorimeters, etc) of the final states. Of
course more bunches means more frequent interactions, requiring
faster read-out electronics, something which however can be achieved in
the future. From the accelerator viewpoint, however, an increased
number of bunches leads to an increase in the electron-cloud
effect. Synchrotron radiation and halo protons hitting the beampipe
wall will extract electrons from it, and these electrons, accelerated
in the field of the passing-by bunches and hitting themselves the pipe
walls, generate a chain reaction where more and more electrons are
released. The energy generated by their interaction
with the beampipe can increase the temperature of the magnets, leading
to a quench, and their presence can furthermore interfere with the main beam
causing its disruption. The size of these effects grows rapidly with
the bunch frequency. Detailed studies of this effect, in view of
the available cooling power and of the beam dynamics, have recently
concluded that previous plans to operate at 12.5~ns are unsustainable,
and all upgrade schemes are now relying on either the nominal bunch
structure, or possibly an increase to a 50~ns spacing, as discussed later.

\subsubsection{Reducing $\beta^*$}
While the overall behaviour of $\beta(s)$ around the full ring is
constrained by global stability requirements, like the value of the
tune, its value at a specific
point depends on the local machine optics. Proper focusing
magnets around the IR can therefore reduce $\beta^*$, thus increasing the
luminosity according to eq.~\ref{eq:lum}. Beam dynamics demands
however a price to pay for a beam
squeeze at the IP: a greater 
growth of the beam size before and after the
minimum of $\beta^*$. The internal aperture of the quadruple magnets
surrounding the IR must match this growth, to prevent the beam hitting
their inner surface. An increase in the quadrupole aperture, on the
other hand, requires a larger field, in order to maintain the
constant field gradient necessary to focus the beam. Alternatively,
one should increase the length of the quadrupoles, so that
the overall bending stays constant. In both cases, the present
focusing quadrupoles surrounding the IRs of the experiments need to be
rebuilt. Notice also that, due to the geometric factor
$F(\phi)$, the reduction in $\beta^*$, with a constant crossing angle,
does not lead to a linear
increase in luminosity.

\subsection{The LHC luminosity upgrade phases} 
The overall goal of the upgrade~\cite{evansAT} is to increase the integrated
luminosity accumulated by the experiments, maintaining their ability
to collect good quality data~\cite{nessiAT}. 
The pursuit of the highest
peak luminosity has therefore to be moderated by considerations of overall
efficiency, providing beam lifetimes as long as possible, refill times
as short as possible, maximal operational reliability (i.e. short
maintenance downtimes), and an optimal experimental environment.  

Two main machine parameters characterize the environment within which
 experiments at the LHC operate: the time between two bunch crossings
 in the center of a detector, $\Delta \tau_b$, and the average number
 of pile-up events, i.e. the simultaneous $pp$ collisions that take
 place in each crossing, $N_{int}$. $\Delta \tau_b$ sets the time
 scale for the frequency at which the detector must sample the data,
 and for the triggering and data acquisition processes. The detector
 signals are to be read and temporarily stored in a buffer while
 hardware and software processing analyze the gross features of the
 event to decide whether it is worth storing, in which case it is
 finally assembled and written out to tape. Under nominal operations,
 $\Delta\tau_b = 25$ns, the reading/processing must be accomplished 40
 million times per second. With an average event size of about 1~MB,
 this means processing 40~PB/s, with the goal of extracting the 100
 interesting events, out of 40M, that in one second that can be
 written to storage for the offline analysis.  $N_{int}$ sets the
 scale of the complexity of the event: each additional $pp$ collision
 during a bunch crossing
 contributes to the occupancy of the detector channels, to the
 processing power needed to trigger, to the size of the event as it is
 stored. In the offline analysis, these pile-up events will
 deteriorate the reconstruction of the interesting final states, as
 discussed in more detail below.  A total inelastic $pp$ cross section
 $\sigma_{pp}$ of about 60mb, the nominal luminosity of
 $L$=\lhclum, and the fraction of the ring populated by bunches
 ($f=2808\times 7.5$~m/27~km), 
give $N_{int} = L\times \sigma_{pp} \times \Delta\tau_b /f
 \sim 19$. This grows linearly with the istantaneous luminosity.

The present upgrade plan of the
LHC~\cite{Scandale,zimmermannAT,garobyAT}  is
tailored to fulfill a gradual luminosity increase with a sequence of
steps, tuned to the construction timescale and complexity of each
step, to the need of accommodating the necessary upgrades of the
experiments, and the desire to maximize the integrated luminosity
delivered by the middle and by the end of the next decade.
The effect of these steps is summarized in fig.~\ref{fig:lumupgrade},
and is briefly outlined in the following sections. 

While the timeline shown here reflects the present planning, it is
expected that it will evolve as a function of the actual performance
of the LHC during its first one or two years of operation. Likewise,
it is still early to have an accurate cost appraisal for the completed
project. Nevertheless, current estimates place the cost of the overall
luminosity upgrade at a fraction of the LHC cost, below the 1 billion
swiss franc threshold. The cost of the upgrade of the SPS and the
energy doubling of the LHC, on the other hand, would be
significantly higher and approach the scale of the initial LHC cost.
\begin{figure}
\begin{center}
\resizebox*{12cm}{!}{\includegraphics{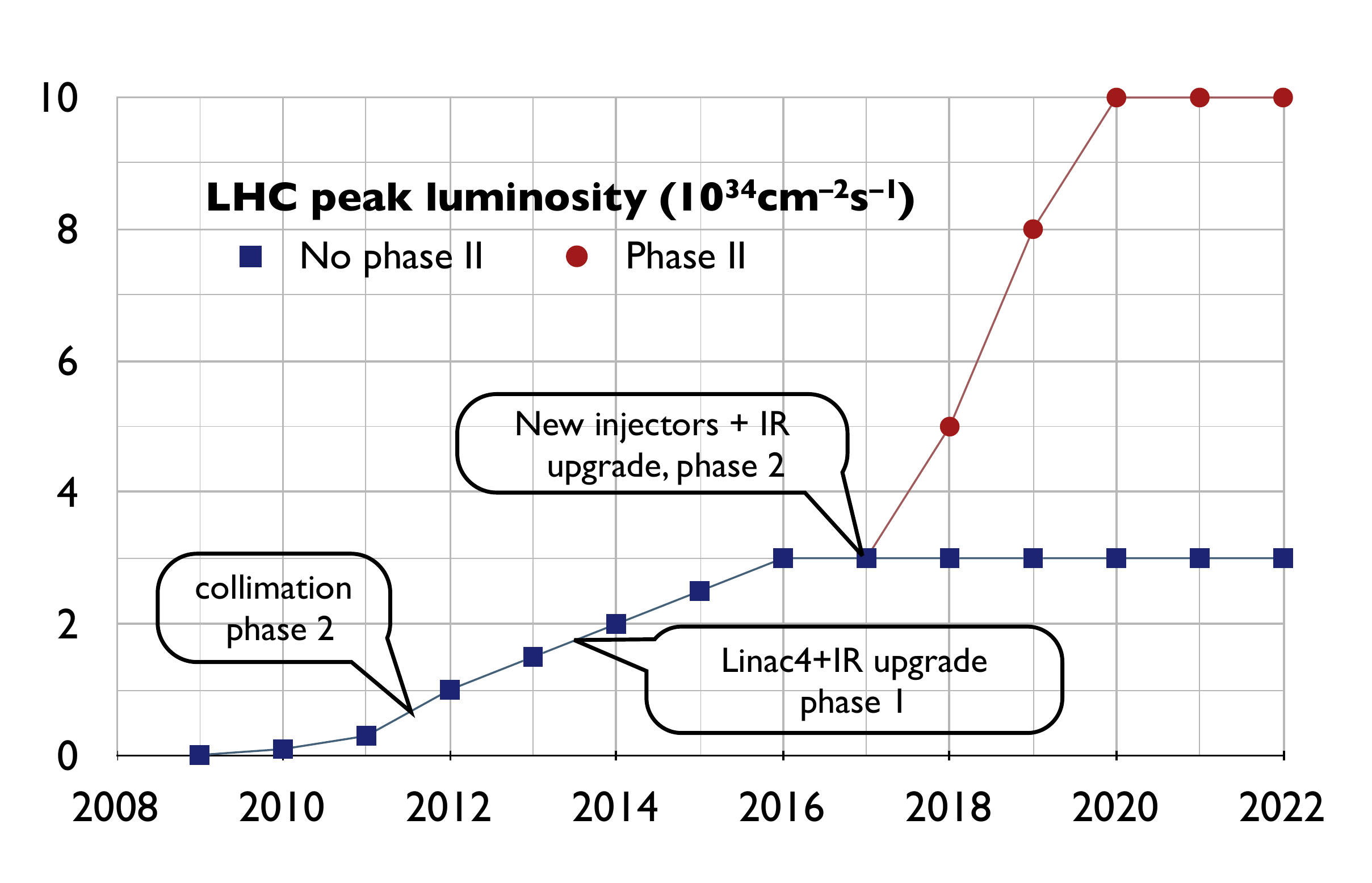}}%
\caption{The peak luminosity profile over the next decade, 
as foreseen by the present upgrade planning.}%
\label{fig:lumupgrade}
\end{center}
\end{figure}

\subsubsection{Achieving nominal luminosity}
While the LHC injector complex is already capable of delivering the
nominal beam brightness, leading to a peak luminosity of \lhclum, it
will take some time and some further LHC hardware before this is
achieved. The main current limitation is due to the beam collimation
system~\cite{assmannAT}. This system of absorbers, inserted in the
beampipe at a close distance from the beam axis and at points far away
from the experiments, ensures that protons in the
beam halos are captured before their orbit leads them to interact with
the beampipe, releasing their energy to the magnets and causing their
quench. It is estimated that the present collimation system allows the
luminosity to reach approximately $0.3\times$\lhclum. An upgrade is
foreseen to allow the nominal luminosity of \lhclum, and beyond. 
The installation
of this system will require a long shut-down of at least 8 months. To
maximize the luminosity integrated during the first few years, and in
consideration of the time required to master operations at such high
luminosity, it is foreseen that this upgrade will take place not before two full
years of running (in fig.~\ref{fig:lumupgrade} this is inserted 
during the 2011-2012 shutdown, labeled as ``collimation phase 2''). 
Once this is done, the LHC luminosity is expected to promptly ramp up
to its nominal value, with an expected yearly integral of about
60~\ifb. 

\subsubsection{The upgrade, phase 1}
This phase will rely on the availability of a new
linear accelerator, the Linac4, to replace the Linac2, and on the
replacement of the IR quadrupoles with greater aperture and greater
field ones. 

The Linac4, whose construction has started and should be
completed by 2014, will raise the injection energy into the PSB from
50 to 160~MeV. The factor of two gain in $\beta\gamma^2$ allows to
double the beam intensity at constant tune shift, providing a better
match to the space-charge limitations of the PSB. The early stages of
the acceleration use an $H^{-}$ beam, whose two electrons will be
eventually stripped off. This step eludes the constraints of Liouville
theorem, and reduces the beam emittance.
Overall, the improved beam quality will allow to increase
$N_b$ from the nominal value of $1.15\times 10^{11}$ to $\sim
1.7\times 10^{11}$, leading to the so-called {\it ultimate} luminosity
of $2.3\times$\lhclum. 

The IRs will need to be upgraded, as discussed in detail
in~\cite{Baglin:2008zz}.  New quadrupoles~\cite{rossiAT}, 
built with the same NbTi
superconducting cable of the LHC bending dipoles, will allow a
reduction of $\beta^*$ to $\sim 30$cm in the ATLAS and CMS IRs. With
this further improvement, the
peak luminosity goal for this phase is of the order of
$3\times$\lhclum, with  a yearly integral of 180~\ifb. 

A long shutdown of about 8 months will be required
for the replacement of the quadrupoles. This is scheduled to take
place between the 2013 and 2014 runs. The shutdown should be
synchronized with the readiness for installation of important
experimental upgrades, such as the trackers.

\subsubsection{The upgrade, phase 2}
The objective of this second phase, also called the sLHC,
is to remove all bottlenecks in the
injector chain, allowing the maximum possible beam brightness to reach
the LHC, and to improve the overall system reliability, renovating
components that, like the PS, are old and require frequent
and time consuming repairs. 
The goal peak luminosity is in the range of \slhclum.
To take full benefit of these beam conditions, a further major
modification of the IRs, beyond phase 1, is required, in order to cope
with the increase in beam-beam interactions and with the deterioration
of the beam lifetime. Furthermore, greater ingenuity is necessary to
limit as much as possible the number of interactions per bunch
crossing, so that the experiments can take full advantage of the
higher interaction rates.
\begin{figure}
\begin{center}
\resizebox*{12cm}{!}{\includegraphics{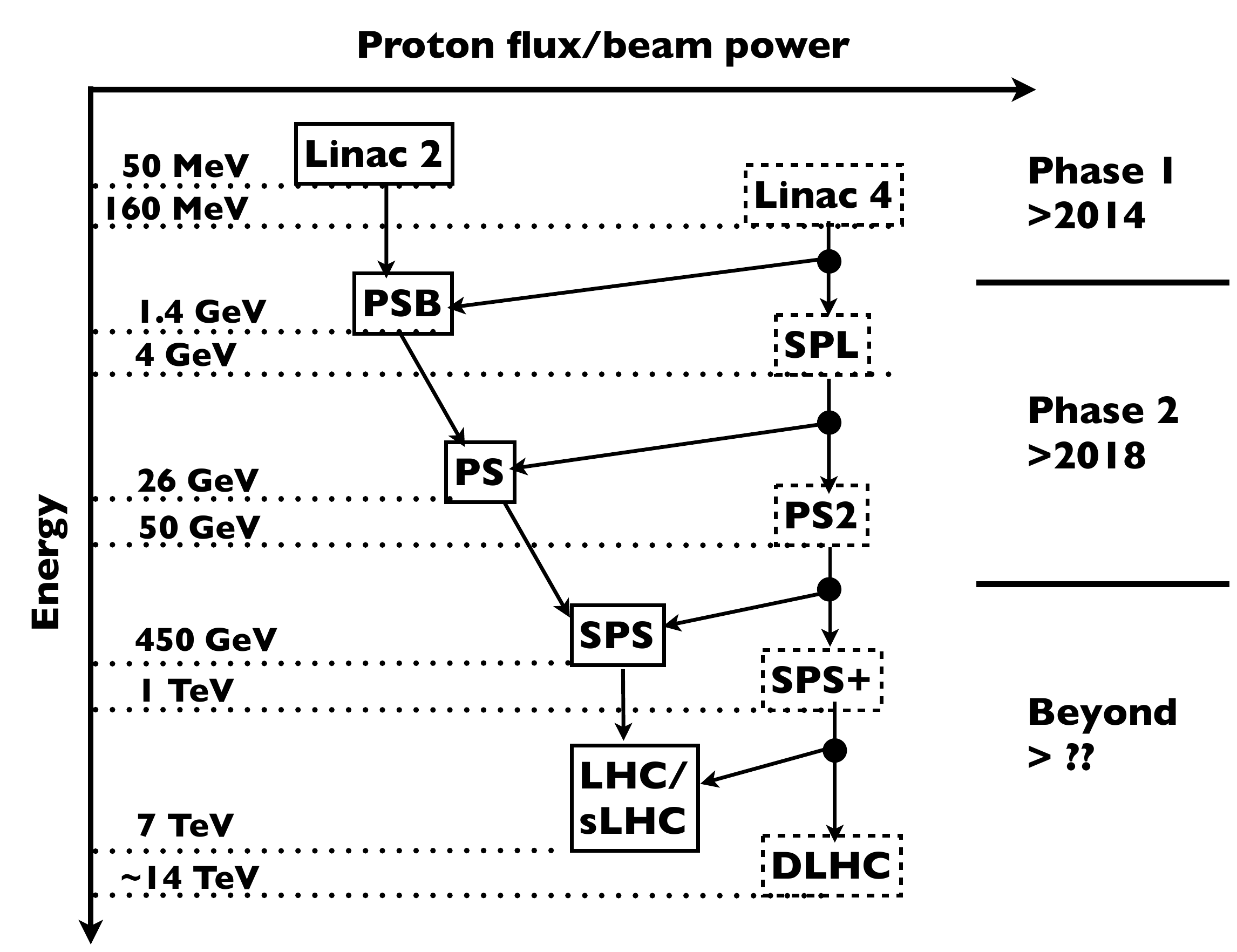}}%
\caption{The evolution of the LHC complex upgrade. The arrows show the
injection sequence, from lower to higher energy accelerators. The
components within dashed boxes on the right-hand side 
are intended to replace the corresponding elements on the left-hand
side. At the completion of phase~2 the chain will link
Linac4, SPL and PS2, then injecting into the SPS and finally the LHC. }%
\label{fig:complexupgrade}
\end{center}
\end{figure}
\paragraph{The upgrade of the accelerators}
The upgrade of the injector complex~\cite{garobyAT} is sketched in
fig.~\ref{fig:complexupgrade}. In addition to the Linac4, which will
already be operating since phase 1, the upgrade calls for a replacement
of both the PSB and the PS. The PSB would be replaced by a
low-power superconducting linear accelerator
(SPL~\cite{Garoby:2005sd}), increasing the injection
energy into the PS from 1.4 to 4 GeV and greatly reducing the filling
time. The increase in output energy of the SPL will
allow to increase also the output energy of the next step in the
chain, where a new synchrotron, the PS2, will replace the PS. The PS2
will deliver protons to the SPS at 50~GeV, well above the 23~GeV
transition energy of the SPS, easying the handling of higher intensities.
Injection into the SPS at 50 GeV will reduce the space-charge tune
spread, to allow the bunch intensity to reach
$N_b\sim3.6\times10^{11}$ for 25~ns bunch separation (and up to
$N_b\sim5.5\times10^{11}$ with 50~ns bunches). 
Higher energy also gives smaller emittance,
and less beam losses at injection. Shorter injection and acceleration
times, finally, reduce the filling time, with a greater operational
efficiency.
\paragraph{The upgrade of the interaction regions}
This great increase in beam intensity needs a complete redesign of the
IRs. Four schemes are presently under
consideration~\cite{Scandale,zimmermannAT}: early separation
(ES) of the beams, full crab crossing (FCC), large Piwinski angle
(LPA) and low emittance (LE).  In the ES scheme additional small
superconducting dipoles are positioned
on either side of the IP, residing within the detectors. This allows to
keep the bunches separated until these final dipoles, reducing the
tune-shift. Small-angle crab cavities, located outside the
quadrupoles, would allow for a total overlap at the collision.  The FCC
scheme solely relies
on crab cavities to maximize the bunch overlap. The LE
scheme provides much thinner bunches, at a cost of a larger geometric
loss. The LPA scheme allows for much more intense beams, requiring a
longer bunch spacing of 50~ns and a larger crossing angle, limiting the
tune shift with a flat beam profile. Long-range beam-beam
interactions need to be screened with compensating wires, to reduce the
tune spread. In this scheme,
lower-intensity bunches separated by 25~ns from the primary ones would
have to be inserted in order to allow collisions in LHCb, whose
position along the ring is out of synch with the collision points of
the 50~ns bunches.

 In general,
more performing quadrupoles, built of Nb$_3$Sn cable, 
will be required, to allow for the reduction of $\beta^*$
envisaged in most schemes, and for the greater aperture needed in the
LPA scheme. R\&D for these new-generation magnets is
underway~\cite{rossiAT}.

\begin{table}
  \tbl{Main parameters of the proposed schemes for the IRs.
 L$_{eff}^{10(5)hr}$ represents the effective luminosity, accounting for
 a LHC refilling time at the end of each store of 10 (5) hours.}
{\begin{tabular}{@{}lcccccc}\toprule
   Parameter  & Nominal & Ultimate & ES & FCC & LE & LPA \\
\colrule
emittance $\epsilon$[$\mu$m] & 3.75 & 3.75 & 3.75 & 3.75 & 1.0 & 3.75 \\
$N_b\,[10^{11}]$ & 1.15 & 1.7 & 1.7 & 1.7 & 1.7 & 4.9 \\
bunch spacing [ns] & 25 & 25 & 25 & 25 & 25 & 50 \\
$\beta^*$ [cm] & 55 & 50 & 8 & 8 & 10 & 25 \\
$\theta_c$ [$\mu$rad] & 285 & 315 & 0 & 0 & 311 & 381 \\
peak L [\lhclum] & 1 & 2.3 & 15.5 & 15.5 & 16.3 & 10.7 \\
$\langle$events/crossing$\rangle$ & 19 & 44 & 294 & 294 & 309 & 403 \\
L lifetime ($\tau_L$[hr]) & 22 & 14 & 2.2 & 2.2 & 2.0 & 4.5 \\
L$_{eff}^{10hr}$[\lhclum] & 0.46 & 0.91 & 2.4 & 2.4 & 2.5 & 2.5 \\
L$_{eff}^{5hr}$[\lhclum] & 0.56 & 1.15 & 3.6 & 3.6 & 3.7 & 3.5 \\
   \botrule
  \end{tabular}}
\label{tab:schemes}
\end{table}
The value of the main parameters
discussed so far, for the four upgrade options and for the pre-sLHC phases, is
shown in table~\ref{tab:schemes}. In particular, notice the important
decrease in the luminosity lifetime with all sLHC schemes. The drop in
luminosity is proportional to the increase in interaction rate, and is
just due to the loss of protons due to the collisions. With a
bunch-bunch crossing each 25~ns, 300 pp collisions at each crossing,
and two active experiments (ATLAS and CMS), a number of the order of
$10^{14}$ p/hr is disappearing, out of a total of about
$5\times 10^{14}$ stored in the initial beam! Under these conditions,
it is critical to reduce as much as possible the time required to
prepare and refill the LHC for a new store. The envisaged turaround
times range from a conservative 10 hours, to a goal of 5 hours; the
impact of this dead time on the effective luminosity is shown in the
last two rows of the table, and the expected luminosity profiles, for
a 5-hour turnaround, is displayed in upper left corner of
fig.~\ref{fig:lumprofile}. 
\begin{figure}
\begin{center}
\begin{minipage}{150mm}
\subfigure[]{
\resizebox*{6cm}{!}{\includegraphics{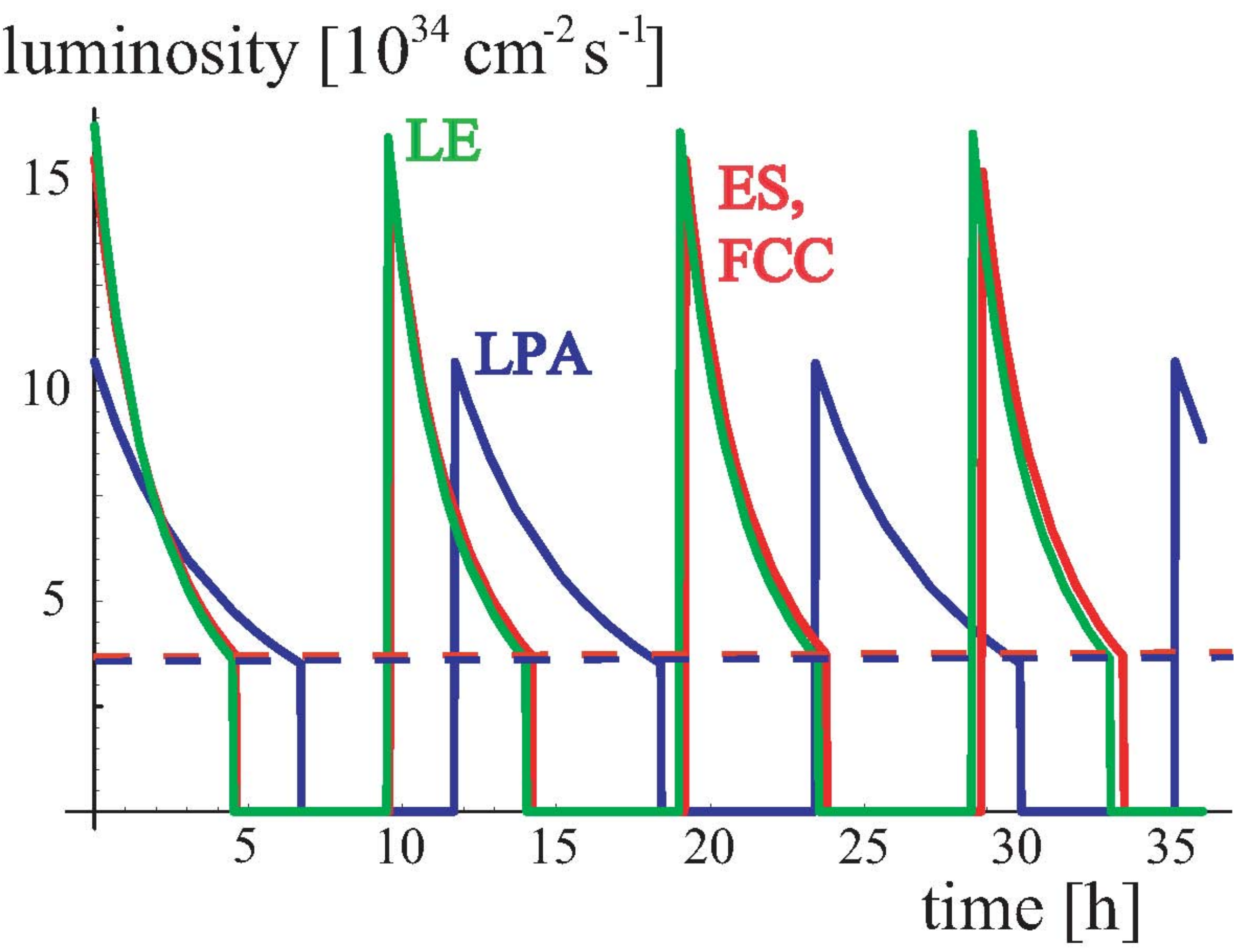}}}%
\hfil
\subfigure[]{
\resizebox*{6cm}{!}{\includegraphics{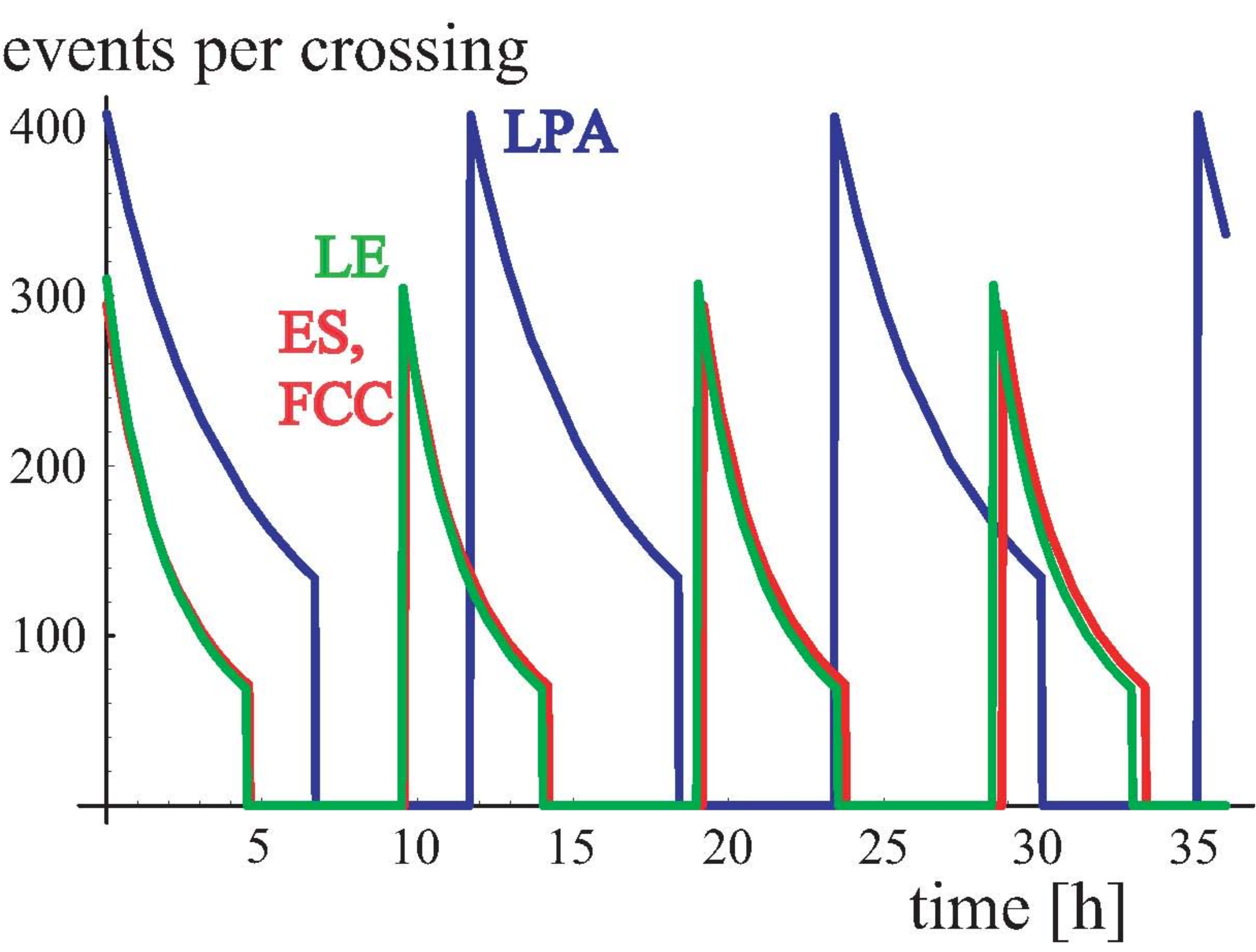}}}%
\\
\subfigure[]{
\resizebox*{6cm}{!}{\includegraphics{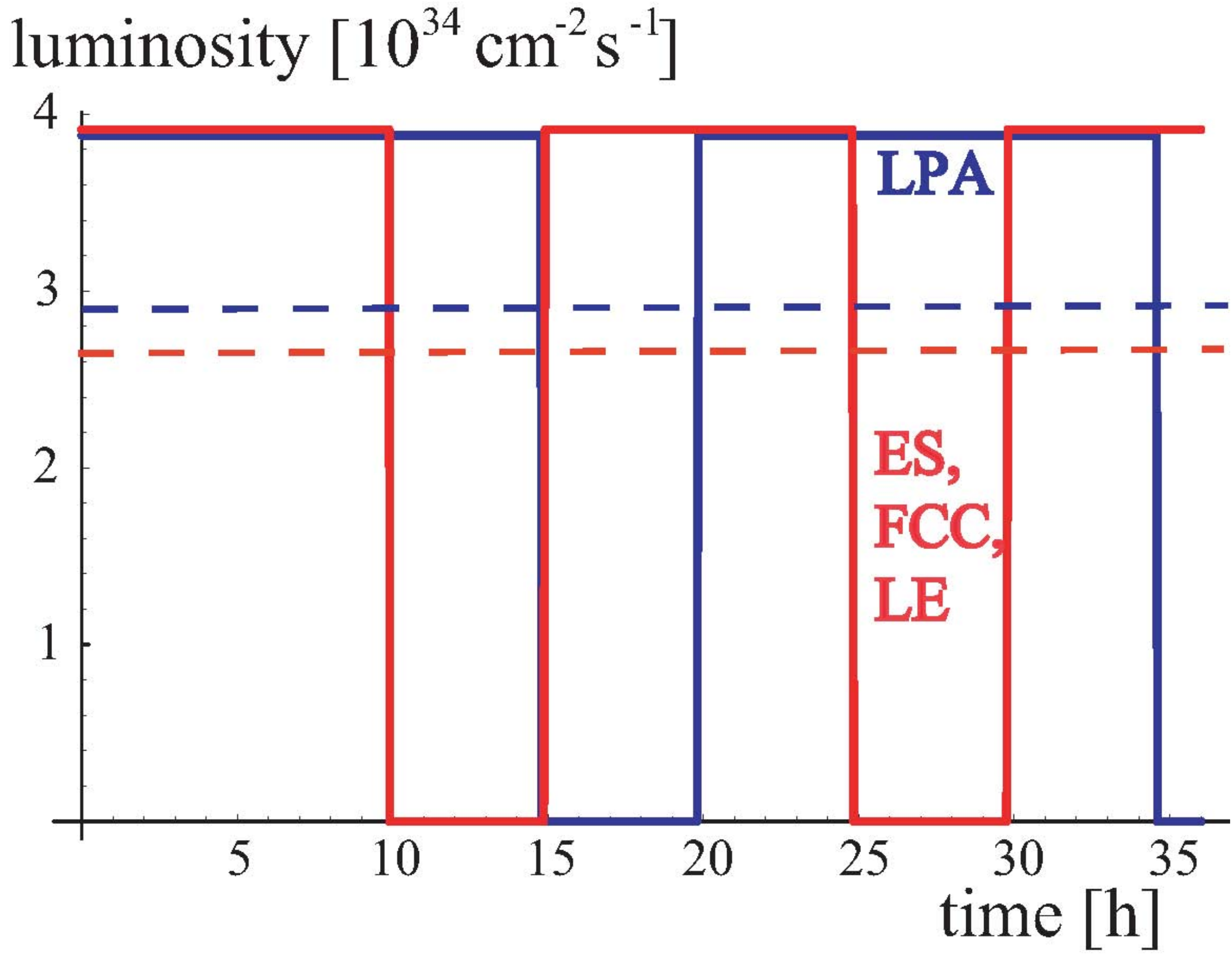}}}%
\hfil
\subfigure[]{
\resizebox*{6cm}{!}{\includegraphics{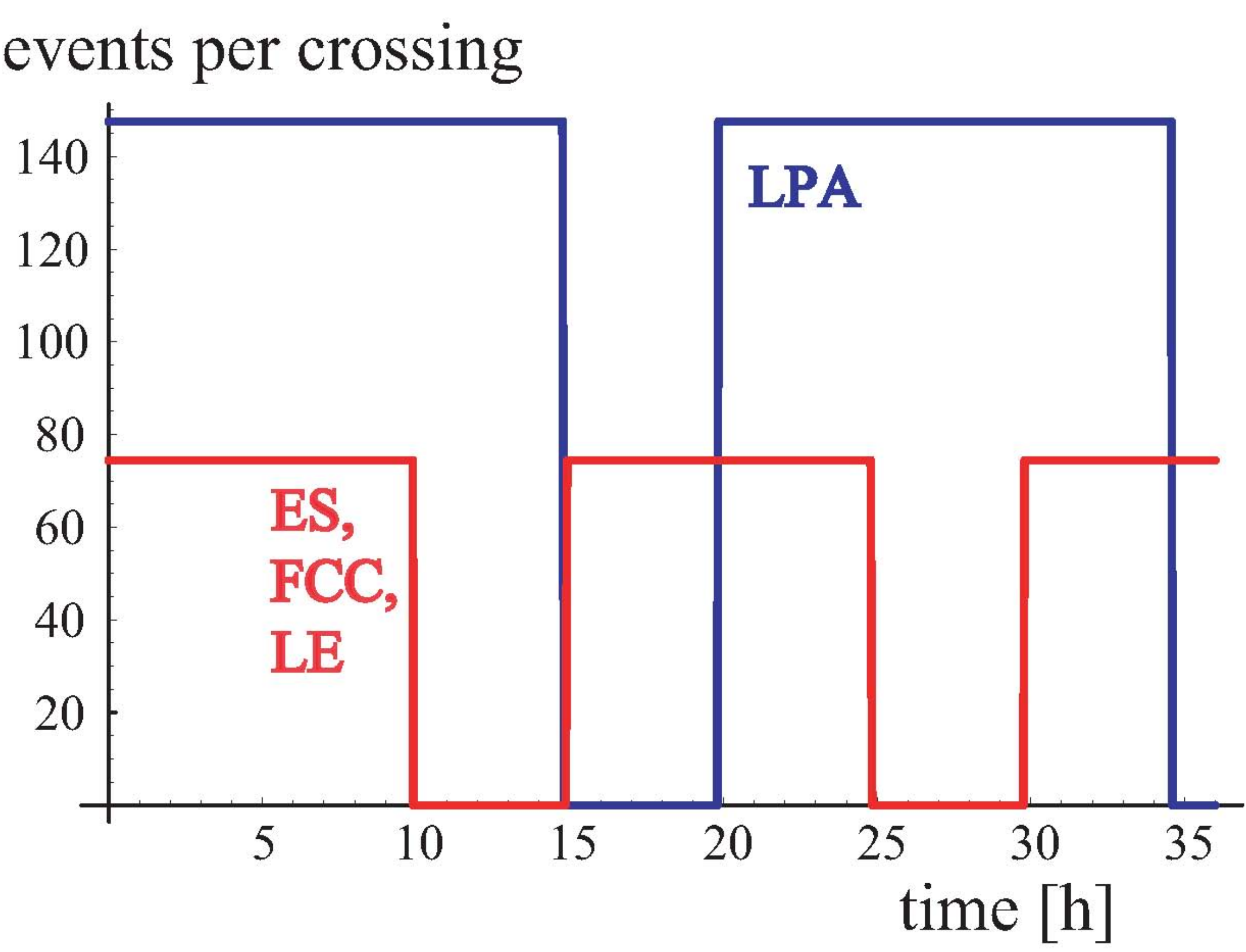}}}%
\caption{Luminosity profile (left side) and number of interactions per
  bunch crossing (right side) for the four IR upgrade options. The
  upper set of plots corresponds to running operations at fixed beam
  parameters, the lower set shows the effect of luminosity leveling.}%
\label{fig:lumprofile}
\end{minipage}
\end{center}
\end{figure}

The upper right plot of fig.~\ref{fig:lumprofile} shows the average
number of expected $pp$ interactions, $N_{int}$, taking place during
each crossing. These numbers, well in excess of 100, should be
compared with $N_{int}\sim 20$ obtained at the nominal luminosity of
\lhclum, a number already putting great strain on the detector
performance. It has been noticed that proper manipulations of the beam
parameters during a single store could level out the luminosity
profile, maintaining the same average luminosity, and providing much
more stable and sustainable conditions for the experiments. The result is
shown in the lower set of plots of
fig.~\ref{fig:lumprofile}. 

Luminosity leveling can be achieved by having a larger value of
$\beta^*$, or a larger crossing/crab angle, early on in the store, and
reducing them as the beam intensity diminuishes during the store. It
is clear that it will provide one of the most powerful tools to
enhance the benefits of the luminosity upgrade, and it is therefore an essential
part of the future planning for the sLHC!

\subsubsection{Beyond phase 2}
The most natural further upgrade of the LHC, beyond phase 2, is the
energy increase, possibly by a factor of at least 2.  The challenge,
magnitude and cost of this effort are significantly larger than the
sLHC~\cite{Bruning:2002yh}.  The precondition for its feasibility is
the development and industrialization of bending dipoles with fields
in the range of 15-20~T~\cite{rossiAT,magnetwshop}.  This can be
achieved in principle with Nb$_3$Sn superconducting cables, such as
those being developed for the phase~2 quadrupoles. Yet higher fields,
around 24~T, or centre-of-mass energies around 40~TeV, could be
achieved with inner-core windings of Bi-2212~\cite{McIntyre:2005ny}.
Aside from the development of the technology needed to produce cables
of the required quality (sufficient current density, strain and
radiation tolerance, etc.)  and length (multi-km!), a major difficulty
of these dipoles will be the management of the immense magnetic forces,
acting both on the overall structure of the magnets, and on their
internal components. The timescale for these developments and for more
conclusive statements about the technological feasibility and cost of
a DLHC (the {\it Double}-LHC) is estimated
to be no less than 10 years. It has to be added that the DLHC requires
also an energy upgrade of the SPS, the SPS+, to boost the injection energy to
1~TeV.

\section{The experimental upgrades}
Three key considerations define the needs of the detector upgrades on
the way towards the sLHC and beyond~\cite{nessiAT}: {\it (i)} some
components will need to be replaced due to the damage caused by
radiation even before the start of the sLHC; {\it (ii)} some
components will not be operable in the harsh high-luminosity
environment of the sLHC; {\it (iii)} the performance of some of the
existing components will not be adequate, at the sLHC, to fulfill the
physics needs.  These issues will be reviewed in this section,
starting from the physics requirements.

\subsection{Physics performance}
\label{sec:physperf}
The criteria for the physics performance of the experiments are driven
by the nature of the observables that will be of interest at the
sLHC. In turn, these are defined by the research programme that will
emerge after the first few years of data taking, once the nature of
the new phenomena of interest will be more clearly defined. 

As mentioned above, 
the main impact of running at high luminosity will be the increased
number of pile-up events, namely the additional $pp$ interactions
taking place during a single bunch crossing. 
\begin{figure}
\begin{center}
\begin{minipage}{150mm}
\subfigure[]{
\resizebox*{7cm}{!}{\includegraphics{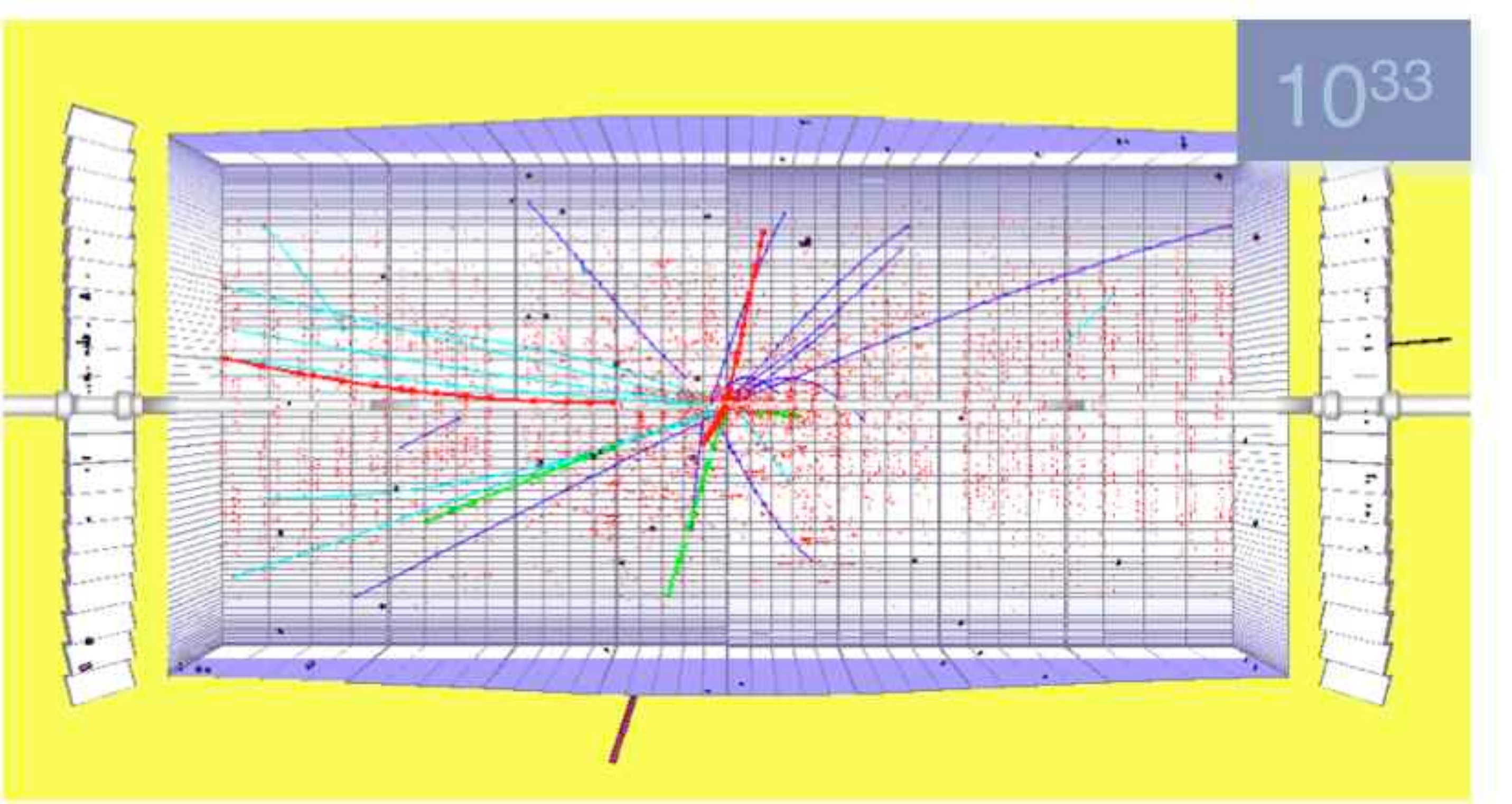}}}%
\hfil
\subfigure[]{
\resizebox*{7cm}{!}{\includegraphics{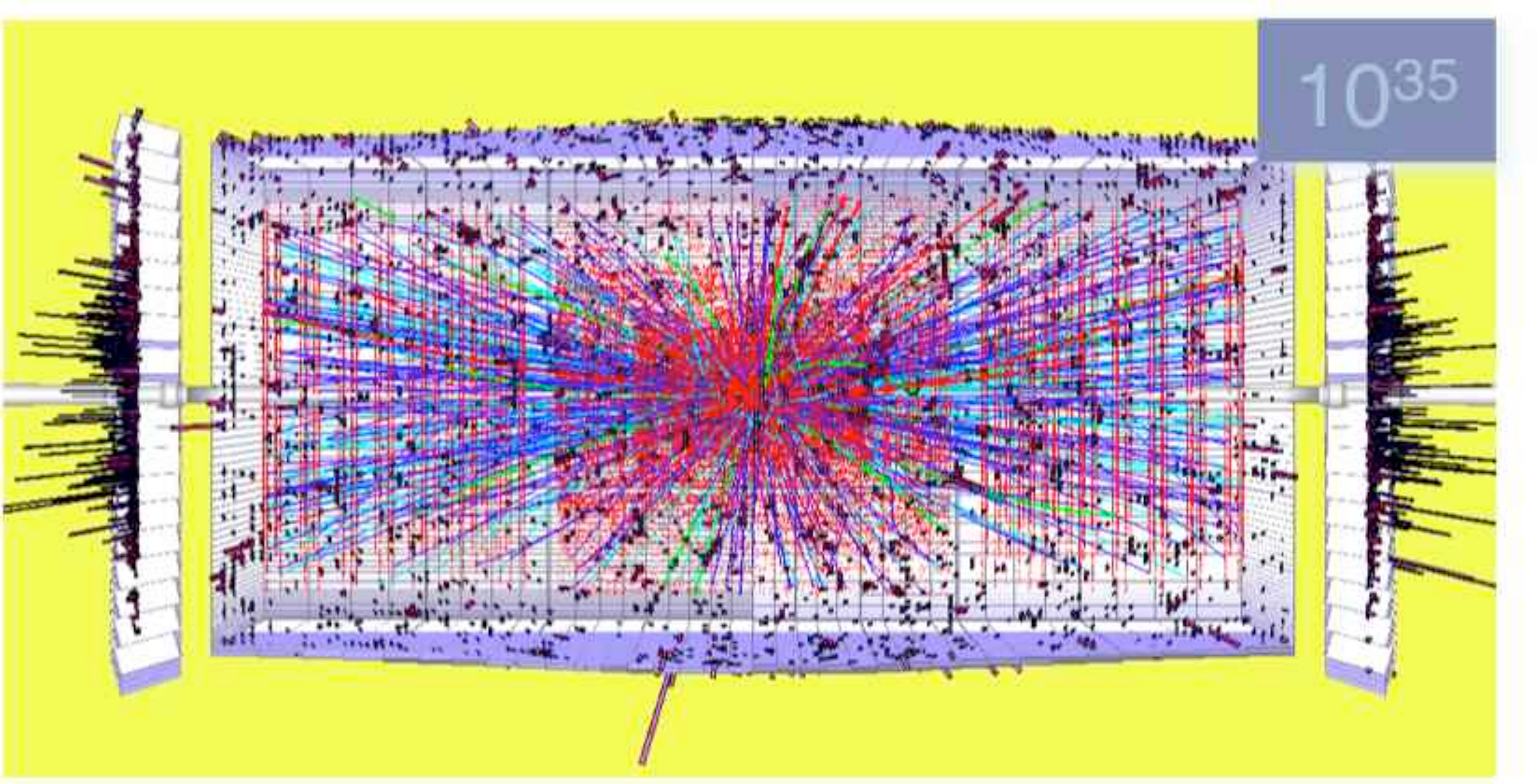}}}%
\caption{The same event, as seen at \lowlum\ (left) and at \slhclum,
  with the inclusion of O(100) additional $pp$
  interactions during the same bunch crossing.}
\label{fig:pileup}
\end{minipage}
\end{center}
\end{figure}
A dramatic picture of what this entails is shown in
fig.~\ref{fig:pileup}, where the same event appears displayed in the
presence of the low pile-up of collisions at \lowlum, and of the
many hundreds of additional events occurring at \slhclum.

The impact of these pile-up events will be twofold. On one side, many
more tracks will be present. This will greatly increase the number of
hits in the tracking detectors, especially at small radius, where
low-momentum particles curling up in the magnetic field cross the
trackers many times. With the increase in occupancy, reconstructing
tracks becomes more and more difficult. The chances to wrongly assign
hits will increase the number of fake tracks, and this extra noise may
prevent the reconstruction of good tracks. The ability to reconstruct
displaced vertices will also deteriorate, with a reduced efficiency to
tag $b$ quarks and $\tau$ leptons, and a larger rate of fake
tags. More hits will also mean much more computing time required to
perform the pattern recognition, affecting both the performance of
tracking triggers, and of the offline analysis. Most of these issues
can be addressed by increasing the granularity of the detectors,
e.g. through a more extended use of pixels even at large radii. This
would reduce the occupancy, avoid overlaps of signals, and improve the
pattern recognition. The penalty is a large increase in the number of
electronics channels, extra heat-load to be removed, and likely an
increase in the amount of detector material, which deteriorates the
momentum and energy measurements because of the increased interactions
of the primary particles.

The other consequence of greater pile-up is the presence of much more
energy within the cones used to reconstruct jets. The extended
transverse size of a jet is determined by physics, and an increased
granularity of the calorimeters, contrary to the case of the trackers,
would not help. The jets from the decay of a top quark, for example,
would collect a large amount of spurious energy from the pile-up
events. Even if their average contribution can be subtracted,
event-by-event fluctuations cannot be disentangled, causing a
significant deterioration in the jet energy resolution.  The
reconstruction of a sharp invariant-mass peak in a two-jet final state
would therefore be harder, and the significance of the peak signal on
top of a large continuum background reduced. Furthermore, electron
identification and trigger efficiency would deteriorate, due to the
larger amount of energy surrounding an otherwise isolated electron,
which will reduce the effectiveness of the usual isolation criteria.
And, last but not least, the large number of additional events can
contribute both to the presence of extra objects (such as jets or
leptons), and of missing transverse energy, due to energy fluctuations
and to undetected large-transverse-momentum particles emitted at small angle.
The additional central jets could jeopardize the use of jet-vetoing in
the study of Higgs production by vector-boson fusion, as discussed in
section~\ref{sec:strongew}, and the high
rates of forward energy could likewise compromise the forward jet
tagging required by these same studies. 

The effect of the above considerations can be qualitatively assessed
by looking at a possible reference process, the production and decay
of a squark-gluino pair sketched in fig.~\ref{fig:susyFS}.
\begin{figure}
\begin{center}
\resizebox*{8cm}{!}{\includegraphics{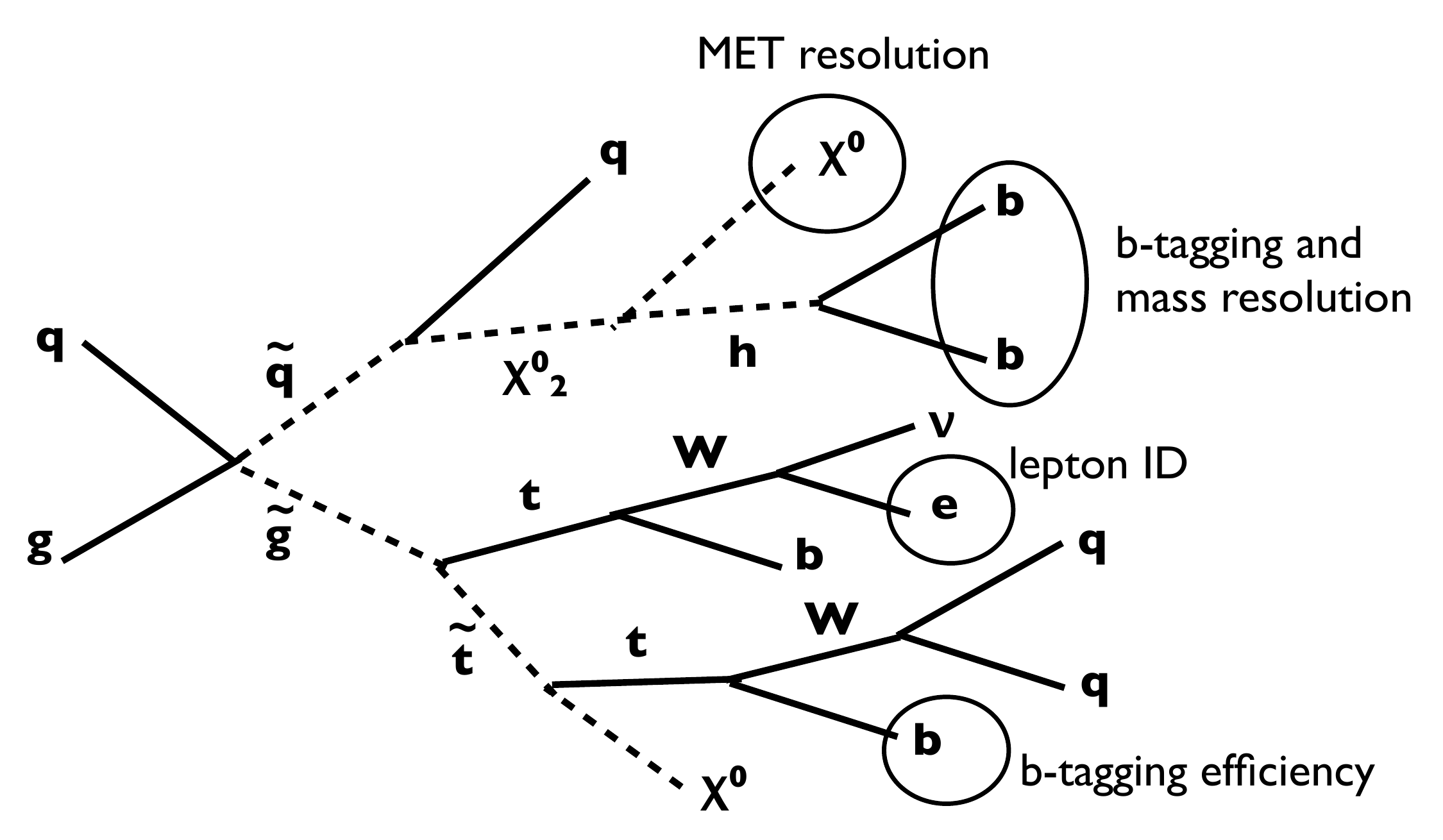}}%
\caption{Example of final states from quark-gluino production in a
  supersymmetric model.}%
\label{fig:susyFS}
\end{center}
\end{figure}
With four $b$-jets and one electron, a mere 20\% loss of
identification efficiency per object would cause a 2.5 loss in
statistics. A 20\% broadening of the mass resolution for the two
reconstructible $W$ and $h^0$ mass peaks would require a further
factor of 1.5 increase in statistics to maintain the same
peak significance. Overall, and neglecting the worsening of missing
transverse energy resolution, this means a net factor of 4 loss in
statistical power relative to a lower-luminosity performance. The
bottom line is that only by keeping the detector performance constant
 can one take full advantage of the promised luminosity increase of the
sLHC! The luminosity leveling discussed above, when seen from this
perspective, acquires a dominant role in any planning.

\subsection{The detectors' upgrade path}
\subsubsection{The transition to phase 1}
Even before worrying about physics performance, the experiments will
have to worry about the operability of their detectors.  By the end of
the nominal luminosity running, towards 2013, the experiments are
expected to have been exposed to approximately 200~\ifb\ of
collisions. At this stage, radiation doses will have compromised the
performance of the innermost layers of the silicon tracking
devices. ATLAS and CMS therefore plan their replacement. ATLAS will
add a new pixel layer, built around a new beampipe, sliding inside the
previous pixel detector. CMS will replace the
entire pixel detector with a new, 4-layers ones.

In addition, both experiments will complete the coverage of the muon
systems, including presently staged new elements, necessary for the
high rates of phase 1. Trigger and data acquisition systems will also
continuosly evolve through the years up to phase 1, to cope with the
yearly rate increases.

The time required for installation of the phase~1 upgrades is of the
order of 8 months, thus consistent with the shut down required for the
IR modifications foreseen by the accelerator plans. Readiness to match
the accelerator schedule of Winter 2013-14 requires that the planning
for these upgrades be completed by early 2010.

In parallel with the ATLAS and CMS preparations for very-high
luminosity, the LHCb experiment is also working towards higher rates.
LHCb can only use a fraction of the full LHC luminosity; at the
beginning, this will be $2\times
10^{32}\ \mathrm{cm}^{-2}\mathrm{s}^{-1}$, and an upgrade programme is
underway to enable LHCb to collect data at $2\times
10^{33}\ \mathrm{cm}^{-2}\mathrm{s}^{-1}$. Since this luminosity still
remains a fraction of the delivered luminosity, both in phase 1 and
phase 2, collisions at LHCb will not significantly impact the
beam lifetime. The main detector improvements 
include new electronics to read out at 40MHz, an upgrade of the
tracking and RICH detectors, and the ability to increase by a factor
of 10 the event output rate, to 20~kHz.

\subsubsection{The transition to phase 2}
In view of the four options currently under consideration for the
accelerator phase~2 upgrade, and of the relative impact on detector
performance, the preference of the experiments goes to the
full-crab-cavity scheme with luminosity leveling. In comparison, 
the LPA scheme has a
pile-up twice as large as that of the other schemes, while the ES scheme
forces the presence of accelerator elements inside the detectors,
compromising, among other things, the forward jet-tagging performance.
Given the still uncertain outcome of the study of the accelerator
options, the experiments are nevetheless exploring upgrade options
capable of getting the best out of each possible running scenario. In
addition to a vigorous R\&D effort on many fronts, new software tools
are also being developed to reliably model the performance in presence
of up to 400 pile-up events.

Some of the key issues for these upgrades are outlined below. A more
complete discussion can be found
in~\cite{Gianotti:2005fm,nessiAT}. 
\paragraph{Tracking}
The current inner trackers of ATLAS and CMS will only survive approximately
700~\ifb, and can operate at best not above \timeslhclum{3}.
New tracking detectors, likely exploiting new technologies such as the
use of diamond, will
therefore be needed by the start of the sLHC.

The detector granularity will be increased to keep occupancy at the 1-2\%
level for an efficient pattern recognition. This can be achieved by
reducing pixel size, strip dimensions for silicon counters, and by adding
more detector layers to increase the number of precision points per
track. This should be done with the constraint of handling the
increased number of channels, the relative heat-load, and trying to
preserve or reduce the material amount, to minimize the interactions
of photons and electrons as they cross the trackers. Several different
technology options and layouts are under consideration.
\paragraph{Muons}
The muon systems of ATLAS and CMS are built with a large safety
factor, needed to accommodate the still unknown amount of backgrounds
expected at the LHC. If the current estimates of backgrounds are
correct, they could in fact maintain their performance even at the
sLHC, at least in the central rapidity region. Options to reduce the
background levels include, for ATLAS, a replacement of the beampipe
with a more transparent one in beryllium, along the full 50m of the
detector. A replacement for the most forward components, nevertheless,
is being considered.
\paragraph{Calorimetry}
The central calorimeters are the most massive components of the
detectors. Their replacement is not an option. Fortunately this is not
required, since, as mentioned earlier, no improvement in their
granularity or energy resolution can compensate for the higher pile-up
environment. The electronics and power supplies will nevertheless be
changed, due to the radiation damage and to improve the flexibility of
the triggers.

The high radiation doses and heat deposition, on the other hand, will
require a major rework on the ATLAS forward liquid-argon
calorimeters. This may call for a replacement of the technology, or
for better active shiedling.  The effect on muon backgrounds of the
presence of new large-aperture quadrupoles in the final focus, and
their shielding, will also need to be reviewed.
\paragraph{Trigger and data acquisition}
 The need is to keep trigger-accept rates constant at each
 level. Relative to the current systems, this means rejecting 10 times
 as many events, and writing 10 times as many bytes, due to the
 increased size of the events in presence of large pile-up.  More
 efficient triggers need to be designed, capable in particular of
 sorting through the massive upfront data flow. The emphasis will
 therefore be on the first trigger levels, where one may need to
 incorporate tracking information, to supplement the reduced rejection
 power of muon and calorimeter triggers, and to maintain acceptable
 efficiency and purity for electrons, affected by the degradation of
 isolation criteria at \slhclum.
\paragraph{General remarks}
The upgrade of the LHC experiments will require major R\&D and
construction work, with a likely timeline of at least 6-7 years for
construction and integration.  The planning has to assume the worse
possible scenarios in terms of pile-up and radiation environment.
While the financial green light for this new enterprise will probably
take a few years and will be tuned to the first LHC discoveries, the
detector community has to act now, preparing technology, making
choices, testing prototypes and going deeply in the engineering
design.

The definition of the scope of the phase 2 upgrades will develop in
parallel with the planning of the IRs upgrades and with the extensive
R\&D programme underway. It is anticipated to converge towards
technical design reports by
2012, together with the complete definition and approval of the
accelerator project.

The goal is to keep the length of the required shutdown to no more than 18
months, to limit to 1 year the loss of beam, and consistent with the
time required in the LHC planning for the IRs changes.

\section{Outlook and conclusions}
Whatever is discovered at the LHC, it will define the research direction
for the future. For example, if supersymmetry is seen, the key problem
of the field will become to understand the origin of supersymmetry
breaking, like today we are confronted with the problem of EWSB. The
first step in this direction will be the measurement of the spectra,
mixings and couplings of the new particles. In the case of the
Standard Model, the complete determination of its parameters has taken
about 30 years, and it would be na\"ive to expect a shorter time scale
for the full exploration of what will be revealed by the LHC.

Hadron accelerators can enjoy great scientific longevity, as shown by
the Fermilab's Tevatron. Its first physics data were taken in 1987;
the first major discovery, the top quark, came in 1994; crucial
results such as the measurement of the $W$ mass and the first evidence
of CP violation in the $B^0$ system arrived few years later.  After a
5-year upgrade shut-down, the CDF and D0 experiments continued their
discovery path, with the first observation of $B_s$ mixing in 2006, an
accurate measurement of $D^0-\bar{D}^0$ mixing in 2007, a
determination of the top mass with accuracy below 1\% in 2008, and
finally reaching, more than 20 years after start-up, sensitivity to
the existence of a Standard Model Higgs boson below~$\sim$~170 GeV.
There is no reason to believe that the LHC should not
deliver significant discoveries and measurements for at least as long.

The complete exploration of the new landscapes unveiled by the LHC
will certainly require additional elements in the experimental
programme, in parallel and beyond the
(s)LHC~\cite{Akesson:2006we,Mangano:2006jp}. In addition to the
development of high-energy lepton
colliders~\cite{Djouadi:2007ik,Accomando:2004sz}, which will improve
the measurement accuracy and sensitivity of the LHC, this programme
includes lower-energy but high-intensity experiments, focused on the
flavour sector of particle physics, notably in the areas of neutrino
physics~\cite{Mohapatra:2005wg}, rare $K$ decays and $B$
physics~\cite{Buchalla:2008jp}, flavour-violation in the charged
lepton sector and electric dipole moments~\cite{Raidal:2008jk}.  The
possibility to address several of these lower-energy programmes with
experiments driven by the beams of the renewed LHC injector
complex~\cite{Blondel:2006vx,Benedikt:2006qf} is an added value to an
upgrade of the LHC, which adds significant physics motivations to its
realization.

While several efforts are underway to define plans for future
accelerators, it is necessary and timely that the full potential of
the LHC be considered and evaluated. The LHC accelerator complex
provides a unique infrastructure, whose upgrades might turn out to be
the most effective way to extend the knowledge about particle physics
in the decades to come. The cost of these upgrades, and their
technological challenge, might turn out to be as demanding as those of
competing projects. On the other hand, the exploratory potential of a
high-energy hadronic collider is unchallanged, and all efforts should
be made to ensure that no stone is left unturned when evaluating how
far the LHC can take us in unveiling the ultimate secrets of
nature. In addition to the luminosity upgrade, which was the focus of
this review, this path should include the consideration of an energy
upgrade, leading to at least a doubling of the center of mass energy.
The current technology provides a natural time ordering to these two
upgrades, with the luminosity one being closer in reach.  The path
towards the luminosity upgrade of the LHC is determined by ipmortant
compromises, characterizing the impact that each technological choice
on the accelerator side has on the detector performance, and
viceversa. The weight to assign to each element depends on the
specific focus of the physics programme, something that may only
become entirely clear after the first few years of LHC operations. A
target the whole community is eagerly waiting for.

\section*{Acknowledgements}
To prepare this review I benefited from the input, implicit or
explicit, of many colleagues with whom I discussed these issues over
the years, as well as of the lecturers of the recent CERN Academic
Training course on {\it Scenarios and Technological Challenges for a
  LHC Luminosity Upgrade}, wisely planned and well organized
 by S. Gobba and J.-P.~Koutchouk.
A likely incomplete list of those I should like to thank includes:
R. Assmann, J. Ellis, L. Evans, R. Garoby, F. Gianotti, N. Hessey,
J. Nash, M. Nessi, G. Rolandi, L. Rossi, F. Ruggiero, W. Scandale,
J. Tuckmantel, J. Virdee, F. Zimmermann.

This work is supported in part by the European Community's Marie- 
Curie Research Training Network HEPTOOLS under contract
MRTN-CT-2006-035505.

\end{document}